\documentclass[english,preprint]{aastex}
\usepackage[T1]{fontenc}
\usepackage[latin9]{inputenc}
\usepackage{textcomp}
\usepackage{amstext}
\usepackage{graphicx}

\makeatletter

\providecommand{\tabularnewline}{\\}

\makeatother

\usepackage{babel}
\usepackage[caption=false]{subfig}

\begin{document}

\title{X-ray Properties of Hard X-ray Selected AGN}

\author{Andrew C. Liebmann}

\affil{Department of Physics, Montana State University, Bozeman, MT 59717,
USA \\
 email: liebmann@physics.montana.edu}

 \author{Andrew C. Fabian}
\affil{Instiute of Astronomy, University of Cambridge, Madingley Road, Cambridge CB3 OHA,
 UK \\
 email:acf@ast.cam.ac.uk}

 \and

 \author{Sachiko Tsuruta}

\affil{Department of Physics, Montana State University, Bozeman, MT 59717,
USA \\
email: uphst@gemini.oscs.montana.edu}


\begin{abstract}
Using the latest 70 month \textit{Swift}-BAT catalog we examined hard X-ray selected Seyfert I galaxies which are relatively little known and little studied, and yet potentially promising to test the ionized relativistic reflection model. From this list we chose 13 sources which have been observed by \textit{XMM-Newton} for less than 20 ks, in order to explore the broad band soft to hard X-ray properties with the analysis of combined \textit{XMM-Newton} and \textit{Swift} data.
Out of these we found seven sources which exhibit potentially promising features of the relativistic disc reflection, such as a strong soft excess, a large Compton hump and/or a broadened Fe line. {\it Longer observations} of four of these sources with the currently operating satellite missions, such as \textit{Suzaku}, \textit{XMM-Newton} and \textit{NuStar} and two others by such future missions as \textit{ASTRO-H}, will be {\it invaluable}, in order to better understand the relativistic disc reflection closest to the central black hole and constrain such important effects of strong gravity as the black hole spin.
\end{abstract}


\section{Introduction}

Several models have been proposed for high energy radiation from accretion-powered active galactic nuclei (AGN). In one kind of reflection models, the ionized relativistic reflection (hereafter referred to as IRR for convenience), the reflection takes place very close to the event horizon of the central black hole so that both special and general relativistic effects play major roles (e.g., \citealt{Miniutti:2004lr, Miniutti:2007bh}). Therefore, by detecting and studying the IRR by satellite observations we can test strong gravity, for instance, by giving constraints on the spin of the central black hole.

Black hole spin can be measured by determining the radius of the innermost stable circular orbit (ISCO) of the accretion disc around the hole. This radius can be found through the studies of the emission reflected from the ISCO region of the optically thick disc (reflection dominated component - hereafter RDC) when it is irradiated by the harder X-ray power law-like continuum component (hereafter PLC) emitted in the corona. The RDC contains a combination of backscattered continuum emission and various atomic line features. The most prominent of these features is very often the Fe K$\alpha$ emission line at around 6.4-6.97 keV (depending on its ionization state). This line is blurred and broadened by  strong relativistic and Doppler effects due to the location very close to the black hole (\citealt{Fabian:1989uq, Laor:1991lr}).

These effects can produce broad and skewed observed profile. If the amount of relativistic blurring can be constrained we can estimate the ISCO radius, and hence the black hole spin. This IRR-like broadened Fe line feature has been observed many times already (e.g., \citealt{Tanaka:1995oq, Nandra:1999kx, Nandra:2007jk, Fabian:2009fk, Walton:2013qy}). However, alternative interpretations also have been offered, such as the absorption based models by \citet{Miller:2008ve, Miller:2009qf}, although \citet{Walton:2012fk} presented a strong argument in favor of the IRR interpretation.

The broad Fe K$\alpha$ feature, however, is not the only characteristic feature of the IRR. The RDC can also display prominent soft excess as well as a broad feature peaking at $\sim$ 30 keV, referred to as the Compton hump. There are numerous emission lines from lighter elements at lower energies than Fe K$\alpha$. These lines are also broadened and blurred by the same relativistic effects, which can blend the complex of reflected lines together into a smooth emission feature which can be interpreted as the observed soft excess
\citep{Crummy:2006ly}. The harder X-ray Compton hump is caused by the combined effects of photoelectric absorption of photons at lower energies and Compton down scattering of photons at higher energies. Therefore, the studies of {\it the broadband spectra from soft to hard X-rays}, covering all these possible characteristic IRR features, will be {\it a powerful method to test this model, and hence the effects of strong gravity}, by, e.g., constraining the black hole spin. In addition, the recent detection of lags between the soft and hard X-ray emission demonstrates that the soft excess responds to changes in the hard continuum, consistent with the IRR scenario (e.g., \citealt{Zoghbi:2011uq}).  Reverberation lags of the Fe-K line that scale with the mass of the black hole have also been detected (e.g., \citealt{Zoghbi:2012kx, Kara:2013vn, Kara:2013ys}).

Other possible explanations have been offered (e.g., \citealt{Gierlinski:2004dq, Middleton:2007vn, Done:2012cr, Jin:2012ys, Miller:2010zr}). However, none of these arguments can deny the possibility of the IRR effects (e.g., see \citealt{Walton:2013qy}).

Recently \cite{Walton:2013qy}  chose 25 `bare' AGN (those with little or no intrinsic absorption) observed by Suzaki, for spectral studies to test the IRR model and determine the black hole spin. These authors found some interesting results, such as predominantly rapid black hole spin within the sample chosen.

To further test IRR it is important that {\it the sample of promising candidates be increased}. Therefore in this paper, we adopted a somewhat different approach - by choosing hard X-ray selected AGN from \textit{Swift} which are less known. These sources are generally not as bright as the better known sources already well observed. However, {\it longer observations} by the existing missions such as \textit{Suzaku} and \textit{XMM-Newton}  combined with \textit{NuStar}, as well as those planned for the near future such as \textit{ASTRO-H}, of these targets will {\it prove to be profitable}. Therefore, in this paper we chose these relatively unknown and yet potentially promising targets, in the hope that their longer observations will be carried out in the near future, to further test the IRR interpretation.

Using the accumulated data of the \textit{Swift}'s Bust Alert Telescope (BAT), a survey of the
hard X-ray sky has been created.  Most recently that is the 70 month catalog by Baumgartner et al. (2013).

\cite{Winter:2009zr, Winter:2010ve} studied the X-ray and optical properties of AGN
from 9 month catalog \citep{Tueller:2008qf}.
\cite{Ricci:2011ly} tested a unified cold reflection model on 165 hard selected Seyferts of all classes.
They found all Seyferts had the same average nuclear continuum and optical depth of the Comptonized material.   Seyfert 2s had on average larger reflections than Seyfert 1s
and Narrow Line Seyfert 1s had stepper hard X-ray spectra.

Section \ref{sec:data redu} summarizes the technical aspect, such as data reduction, of short time observations of our selected sources which were already carried out by \textit{XMM-Newton}  and \textit{Swift}. Section \ref{sec:selection} explains our selection process, and Section \ref{sec:spectra}  shows the results of our spectral analysis. Discussion and conclusions are given in Sections \ref{sec:discution} and \ref{sec:conclusion}.

\section{\label{sec:data redu}Observations and Data Reductions}

\textit{XMM-Newton} reformatted telemetry is organized in the Observation
Data Files (ODFs). The ODFs were processed to produce calibration
event lists by using the \textit{XMM-Newton} Science Analysis System
(SAS) version 12.0.1. The event lists of MOS \citep{Turner:2001rr}
and pn \citep{Struder:2001cr} were generated by the task EMPROC and
EPPROC respectively. The events were further cleaned by removing hot,
dead, and flickering pixels. We used X-ray events corresponding to
pattern 0-12 (single, double, triplet, and quadruple pixel events)
for the MOS cameras and pattern 0-4 (single and double) for the pn
camera. Source spectra and light curves were extracted from a circular
region with a radius of 45 to 60 arcsec depending on the brightness of the source.
Due to the short nature of each
of these observations, the data from Reflection Grating Spectrometer
(RGS) \citep{den-Herder:2001nx} were not used for scientific purposes.

Source spectra were extracted using a spatial filter in dispersion/cross-dispersion
coordinates. Subsequently, the first order events were selected by
applying a dispersion/pulse-height filter. Background spectra were
generated using the events that lie outside the source spatial region.

\textit{Swift}-BAT data were averaged over first 70 months of
operation.  We used the eight-band spectra and response matrixes
as published by \cite{Baumgartner:2013qy}.  See \cite{Baumgartner:2013qy} for
details of their data reduction methods.

Spectral fits were performed with the X-ray spectral fitting package
XSPEC version 12.7.1 \citep{Arnaud:1996ud}. All of the source spectra
were grouped such that each bin contains at least 30 counts in order
to perform the $\chi^{2}$ minimization technique. All fit parameters
are given in the source rest frame, and errors are quoted at the 90\%
confidence level for one interesting parameter (i.e., $\Delta\chi^{2}$
= 2.7), unless otherwise stated. The Galactic column density, \textit{N$_{H}$},
towards each source was fixed at the weight average value of the sky
as reported in \citet{Kalberla:2005kx}. Throughout this paper, H$_{0}$
= 70 km s$^{-1}$ Mpc$^{-1}$, $q{}_{0}=0.0$ and $\Lambda_{0}=0.73$
are assumed.

\section{\label{sec:selection}Selection of Target}

In order to choose sources which are suitable as the IRR candidates, our first criteria is that they are bright at hard X-rays since that can be an indication for the hard Compton hump. However, we considered that in addition some other IRR features in lower energy bands, such as a large soft X-ray excess and possibility for a broad Fe line, are just as important. Also it is desirable that they are less contaminated by absorbing clouds and other complications. 

With these constraints in mind, our target selection process began with the \textit{Swift}-Bat catalog, using the most recent version, the 70 month catalog by Baumgartner et al. (2013). Of the
1210 sources in the catalog 711 are identified with AGN. We eliminated
beamed sources and heavily absorbed sources (such as blazars, Seyfert
IIs, radio galaxies and LINERs), since these sources are not relevant for
our current study. Our current aim is to increase the statistics by finding
relatively unknown sources with little previous data and yet promising for
the IRR. Therefore we removed any sources observed with \textit{Suzaku}
and/or \textit{XMM-Newton} for longer than 20 ks.

We found 13 of \textit{Swift}-BAT sources which exhibited these potentially
interesting IRR
features and yet observed with \textit{XMM-Newton} with less than 20 ks.
Table \ref{tab:obs log} shows a list of these target sources,
their basic properties, and an observation log for \textit{XMM-Newton}.
The sources are listed in the order of higher to lower priority.
In determining our priority we had to take into account delicate balance among
various competing factors, but our top emphasis was the evidence for potentially
interesting IRR features of various kinds, which exceeded the negative factor 
from the absorptions in some cases.

\section{\label{sec:spectra}Results: Spectra Analysis}

\subsection{Basic Approach for the Analysis and Results}

We begin by modeling the spectra by fitting an absorbed power law
from 2.0 keV to 10.0 keV but excluded the region of 5.0 to 7.0 keV.
The results are summarized in Table \ref{tab:simple pl}. These models
are then extended to include the whole bands with \textit{XMM-Newton} and \textit{Swift}-BAT.
The results are shown in Figure \ref{fig:ratio plots}. It shows the
data to the model ratio for each of these sources.

We then applied the following model: A power law, a cold reflection
with narrow Fe line (PEXMON) \citep{Magdziarz:1995dp}, and an ionized
relativistic reflection (IRR) (REFLIONX) \citep{Ross:2005qe}. These
components are covered by a Galactic column density, an ionized absorber
(ZIXPC) and an intrinsic hydrogen column density. Some sources did
not require this ionized absorber as the column density was very low and hence
it was not included. The best fit parameters are summarized in Tables
\ref{tab:spectal fits} and \ref{tab:mod parameteres}.
Most of the sources studied here are well fit by this model and they
have well defined parameters. The best fit parameters are summarized
in Table \ref{tab:spectal fits}.

As all of these sources show some minor variation and the \textit{Swift}-BAT is
averaged over 70 months and the \textit{XMM-Newton} is a single short observation,
we do not have an accurate cross-normalization between the two instruments.
To account for this we linked the parameters of the \textit{XMM-Newton} data to
those of \textit{Swift}-BAT data with the normalization constant left as a free parameter.

We note that all of these sources exhibit at least one or more of potentially interesting
IRR features.


\subsection{Individual Source Details}

A brief summery of each source is given, in the order of higher to lower priority.
All hard X-ray fluxes are given for 14 to 195 keV by \cite{Baumgartner:2013qy}.

\textbf{\textsl{LEDA 168563:}} It is the second brightest hard X-ray source
in this {\it Swift}-BAT sample with $6.32\times10^{-11}$ erg cm$^{-2}$ s$^{-1}$ . It is well fit by the continuum of
power law plus IRR model.
In addition to an intrinsic absorber, $N_{H}=1.26_{-0.08}^{+0.08}\times10^{21}$
atom cm$^{\text{-2}}$, a highly ionized absorber was also included
with $log\:\xi=3.05_{-0.12}^{+0.12}$ and $N_{H}=3.63_{-1.35}^{+1.12}\times10^{21}$
atom cm$^{\text{-2}}$. Although the absorption is not serious, this results
in a substantial improvement to the
fit of $\Delta\chi^{2}=29.80$ for 2 additional degrees of freedom.
The spin parameter was also well constrained, with $a=0.62_{-0.02}^{+0.07}$.

\textbf{\textsl{IGR J16558-5203:}} This source is also well fit by the continuum
of power law plus IRR model. The high energy {\it Swift} flux is $4.55\times10^{-11}$ erg cm$^{-2}$ s$^{-1}$.
The effect of the absorption is insignificant, but in addition to an intrinsic absorber
with $N_{H}=9.90_{-1.37}^{+1.36}\times10^{20}$ atom cm$^{\text{-2}}$,
a low ionized absorber was also included, with $log\:\xi=2.74_{-0.36}^{+0.58}$
and $N_{H}<7.96\times10^{20}$atom cm$^{\text{-2}}$. This resulted
in a slight improvement to the fit of $\Delta\chi^{2}=3.50$ for 2
additional degrees of freedom. The spin parameter was also well 
constrained, with $a=0.71_{-0.05}^{+0.06}$.

\textbf{\textsl{SBS 1301+540: }}This source is well fit by only the
power law plus IRR model. Its hard X-ray flux from {\it Swift} is $3.26\times10^{-11}$ erg cm$^{-2}$ s$^{-1}$.
No additional absorption components were needed.
The lack of a need for an absorption model may indicate
that the feature seen near 0.9 keV could be an Fe L line. See Section
\ref{sub:future sources} for further discussion of this source.

\textbf{\textsl{AX J1737.4-2907:}} This is the brightest source in terms of
hard X-ray of this {\it Swift} sample with $11.59\times10^{-11}$ erg cm$^{-2}$ s$^{-1}$.
It is well fit by the continuum of power
law plus IRR model. In addition to an intrinsic absorber with $N_{H}=8.83_{-0.20}^{+0.20}\times10^{21}$
atom cm$^{\text{-2}}$, an ionized absorber was also included with
$log\:\xi=2.17_{-0.07}^{+0.07}$
and $N_{H}=1.26_{-0.11}^{+0.12}\times10^{22}$atom cm$^{\text{-2}}$.
This results in an improvement to the fit of $\Delta\chi^{2}=19.79$
for 2 additional degrees of freedom.
The spin parameter was also well constrained, with $a=0.97_{-0.02}^{+0.071}$.

\textbf{\textsl{IGR J16119-6036:}} The source is well fit by the continuum
of power law plus IRR model. Its hard X-ray {\it Swift} flux, $3.22\times10^{-11}$ 
erg cm$^{-2}$ s$^{-1}$, is somewhat low. However
the absorption is very low although a low intrinsic absorber,
with $N_{H}=4.72_{-1.04}^{+1.01}\times10^{20}$
atom cm$^{\text{-2}}$, was also included.

\textbf{\textsl{Mrk 1501:}} The source is well fit by the continuum of power law
plus IRR model. Its hard X-ray flux from {\it Swift}, $3.14\times10^{-11}$ erg 
cm$^{-2}$ s$^{-1}$, is somewhat low.
The absorption is very low, though a low intrinsic absorber, with $N_{H}=3.38_{-0.77}^{+0.76}\times10^{20}$
atom cm$^{\text{-2}}$, was also included.
The spin parameter was also well constrained, with $a=0.85_{-0.05}^{+0.04}$.

\textbf{\textsl{IRAS 09149-6206:}} This source was not fit by any of our simple 
models used here. Hard X-ray flux from {\it Swift} is $3.18\times10^{-11}$ erg cm$^{-2}$ s$^{-1}$.
A highly complicated absorption contaminates this source. See Section
5.3 for further discussion on this source.

\textbf{\textsl{IRAS 05218-1212:}}  The source has a very flat apparent spectrum

  It is well fit by the power law plus IRR continuum model.
Hard X-ray flux from {\it Swift} is $1.83\times10^{-11}$ erg cm$^{-2}$ s$^{-1}$. In addition to a low intrinsic absorber, with
$N_{H}=6.57_{-0.01}^{+0.02}\times10^{20}$ atom cm$^{\text{-2}}$,
an ionized absorber, which was needed, was also included, with $log\:\xi=2.01_{-0.01}^{+0.01}$
and $N_{H}=17.52_{-0.80}^{+76}\times10^{22}$ atom cm$^{\text{-2}}$.
This results in a large improvement to the fit of $\Delta\chi^{2}=699.23$ for 2 additional
degrees of freedom.

\textbf{\textsl{NVSS 193013+341047:}}  The source has a very flat apparent spectrum. 
The continuum can be moderately modeled by the power law plus IRR model. Hard X-ray flux from {\it Swift} is $2.74\times10^{-11}$ erg cm$^{-2}$ s$^{-1}$.
A large ionized absorber is needed, with 
$log\:\xi=1.92_{-0.02}^{+0.01}$ and $N_{H}=20.22_{-0.26}^{+0.25}\times10^{22}$ atom cm$^{\text{-2}}$, with a low intrinsics absorber with $N_{H}=3.38_{-0.38}^{+0.39}\times10^{21}$ atom cm$^{\text{-2}}$.

\textbf{\textsl{IGR J07597-3842:}} This source is well fit by the continuum
of power law plus IRR model. Hard X-ray flux from {\it Swift} is $5.06\times10^{-11}$ erg cm$^{-2}$ s$^{-1}$.
In addition to an intrinsic absorber, with
$N_{H}=9.38_{-0.92}^{+0.92}\times10^{22}$ atom cm$^{\text{-2}}$,
a highly ionized absorber was also included, with $log\:\xi=3.22_{-0.12}^{+0.12}$
and $N_{H}=1.03_{-0.51}^{+0.53}\times10^{22}$ atom cm$^{\text{-2}}$.
This results in a minor improvement to the fit of $\Delta\chi^{2}=6.68$
for 2 additional degrees of freedom.

\textbf{\textsl{NGC 4138:}}  This source is poorly fit by the continuum of the power law
plus IRR model. Hard X-ray flux from {\it Swift} is $3.00\times10^{-11}$ erg cm$^{-2}$ s$^{-1}$.
The flat spectrum and dip below 3.0 keV implies a large absorption.
An intrinsic absorber, with $N_{H}=2.49_{-0.23}^{+0.26}\times10^{21}$ atom cm$^{\text{-2}}$, and moderately ionized absorber was also included, with $log\:\xi=1.92_{-0.02}^{+0.01}$
and $N_{H}=20.22_{-0.26}^{+0.25}\times10^{22}$ atom cm$^{\text{-2}}$. They were needed to archive
any reasonable fit.

\textbf{\textsl{Fairall 1146:}}  This source is well fit by the continuum
of power law plus IRR model. Hard X-ray flux from {\it Swift} is $2.89\times10^{-11}$ erg cm$^{-2}$ s$^{-1}$.
In addition to a low intrinsic absorber, with
$N_{H}=1.38_{-0.01}^{+0.01}\times10^{20}$ atom cm$^{\text{-2}}$,
an ionized absorber, which was needed for better fits, was also included, with $log\:\xi=2.07_{-0.02}^{+0.02}$
and $N_{H}=46.15_{-2.67}^{+2.41}\times10^{22}$ atom cm$^{\text{-2}}$.
This results in a large improvement to the fit of $\Delta\chi^{2}=700.90$ for 2 additional
degrees of freedom.

\textbf{\textsl{UGC 12138:}} It is the dimmest in hard X-ray flux of this sample, with $1.66\times10^{-11}$ erg cm$^{-2}$ s$^{-1}$.
However, this source is well fit by the continuum
of power law plus IRR model.  In addition to a low intrinsic absorber, with
$N_{H}=1.26_{-0.01}^{+0.01}\times10^{20}$ atom cm$^{\text{-2}}$,
an ionized absorber, which was needed, was also included, with $log\:\xi=1.64_{-0.17}^{+0.018}$
and $N_{H}=1.58_{-0.39}^{+0.36}\times10^{21}$ atom cm$^{\text{-2}}$.

\section{\label{sec:discution}Discussion}

\subsection{\label{sub:general char}General Characterises of Sources}

First of all, we note that all sources, except for AXJ1737.4-2907 and NGC 4138, possess strong soft excesses. In all cases an ionized relativistic reflection (IRR) model was adequate to explain the soft excess.
For AXJ1737.4-2907 and NGC 4138 the apparent lack of a visible strong soft excess is likely due to a large column density in each of their absorption clouds.

Secondly, Fe lines are detected in all sources. Most of them appear to include broad Fe lines which dominate the region in many cases.
A majority of these are centered around 6.4 keV.
However there are a few that is centered around higher energies, at
6.70-6.98 keV, with potentiality highly ionized Fe. One source, AX
J1737.4-2907, shows a line centered at around 5.99 keV. This may be
due to lack of good data quality and may not be real.  A full list of line intensity and their
equivalent widths is given in Table 5.

Several sources have what appears to be a large Compton hump. That might
not be conclusive due to a combination of
low spectral resolution and instrument sensitivity, as well as poor cross calibration between
the 70 months-averaged data of the {\it Swift}-BAT and the short snapshot
 \textit{XMM-Newton} EPIC data.
However what can be seen is that the power
law used to fit the \textit{XMM-Newton} data from 2.0-10.0 is insufficient
to describe the \textit{Swift}-BAT data, see Figure \ref{fig:ratio plots}.
\citet{Baumgartner:2013qy} was able to describe most of these sources
as a power law. However Fairall 1146 is not well fit by a simple power law and has statically unacceptable fits with $\chi_{r}^{2}>2.0$ \citep{Baumgartner:2013qy}. In several cases the slope between 14 keV to 195 keV as reported in \citet{Baumgartner:2013qy} is steeper than
the \textit{XMM-Newton} power law fitted between 2.0 keV to 10 keV.
If we extend the power law from the \textit{XMM-Newton} data to the \textit{Swift}-BAT data, several
sources exhibit a different slope - there is generally excess between
10 and 40 keV for these sources. The simplest interpretation is that is
the result of a Compton hump caused by reflection.
Future observations by better instruments, especially for hard X-rays, on board, e.g.,
\textit{ASTRO-H}, will be important, for they will settle this current uncertainty.

Not every source has this difference in power law slopes. The remainder
of the sources have a very similar power law between 2.0 keV to 10.0
keV and 14.0 keV to 195 keV. Given the presence of a narrow Fe K$\alpha$ line
in many of these sources there is likely a cold distant reflection also.
However due to the limited spectral resolution and large error bars
of the current data we are currently unable to say more.

Due to very short {\it XMM-Newton} observation time, the currently available data were only able
 to constrain the black hole spin parameter in a few cases.
 Spin parameters for LEDA 168563, IGR J16558-5203, AX J1737.4-2907 and Mrk 1501 were able to be constrained with the current data.
 All other sources were not constrained, but future longer observations
by {\it XMM-Newton}, \textit{NuStar} and \textit{ASTRO-H} will be able to constrain this important parameter.

AX J1737.4-2907, IRAS 09149-6206, NGC 4138, Fairall 1146, NVSS 193013+341047, and
IRAS 05218+1212 all have noticeable large intrinsic
absorption. A majority of these are due to what appears to be a simple
intrinsic absorption caused by a thicker neutral hydrogen column density.
However the absorption seen in IRAS 09149-6206 is not caused by such
a simple model. A much more complicated absorption, such as a warm
absorber, is occurring.  If the flat spectra of NGC 4138, NVSS 193013+341047 and
IRAS 05218+1212 are caused by large column densities, they are an order of magnitude
larger than the other sources.

IRAS 09149-6206 and NGC 4138 are not well fit by any of
the simple models used here. In the case of IRAS 09149-6206 the absorption
is too complicated to be fit to a simple model. The combination of
a strong absorber, soft excess, a potentially broad Fe line and large
Compton hump can be difficult to be fit. See Section \ref{sub:other sources}
for further discussion.

NGC 4138, NVSS 193013+341047 and IRAS 05218-1212 all
suffer from a flat spectrum and lower statistics, and they
require large absorption column densities with high ionization parameters.
However, with the exception of NGC 4138, these sources resulted in a reasonable fit.
See Section 5.3 for further discussion.

\subsection{\label{sub:future sources}Potentially Best Sources for Investigation}

For a source to be most suitable as an IRR candidate, it is desirable that
it likely exhibits promising IRR features such as a large soft excess and a broad iron line, as well as being bright at hard X-ray which is a possible indication for the hard Compton hump. Also it is better if it is less contaminated by absorbing clouds.

\bigskip \noindent {\bf 5.2a Best Candidates for Current Missions}

With these criteria in mind, from the above list we have identified four sources that are potentially
promising for farther investigation by currently existing missions, because all of those sources
have strong soft excess, they are bright in the hard band and they have
wider Fe lines. These are LEDA 168563, IGR 16558-5203, SBS 1301+540, and AX J1737.4-2907.
Moreover, except AX J1737.4-2907, they are miniumally affected by absorbers.

Possessing the second brightest hard X-ray flux and the largest
soft excess on the list (see Figure \ref{fig:ratio plots}), LEDA 168563 is a clear target for farther observations. It possesses all the desirable elements needed to test the
IRR model. Additionally this source seems to have a possibly broad Fe
feature and less absorption.
This source bares some resemblance to some IRR candidates from Figure
1 of \citet{Walton:2013qy}, (e.g., 1H 0419-577, Ark 564, etc.). When
fitted with a Gaussian, the line is centered around $6.45_{-0.06}^{+0.05}$
keV with an EW of $38.0_{-18.6}^{+18.6}$ eV.

IGR 16558-5203 does not have as large soft excess as, nor it is as
bright in the hard band, as LEDA 168563, but still has a large hard X-ray flux. More
importantly, however, with negligible
intrinsic absorption and a distinct, strong broad Fe line it still is a very promising source.
The soft excess and Fe line are similar to other promising IRR sources
(e.g., Ark 564, Mrk 335, etc.) seen in Figure 1 of \citet{Walton:2013qy}.
When fitted with a Gaussian the line is centered around $6.49_{-0.06}^{+0.07}$
keV with EW of 115.1 eV.

SBS 1301 is the weakest of the three top candidates, both in terms of soft excess
and hard X-ray flux.
Nevertheless it has important advantages, due to its strong broad Fe
K$\alpha$ line and negligible absorptions. When the Fe
line is fitted by a Gaussian the line is centered around 7.0 keV.
The width is unrestrained, but has a EW of 78.5 eV. This source bares
some resemblance to some IRR candidates from Figure 1 of \citet{Walton:2013qy},
(e.g., Mrk 110, Mrk 359, etc.)  Between 0.7-1.0 keV there is a feature
that could be either an O VII edge due to a warm absorber or a Fe
L line, though the current data is insufficient to tell the difference.
Due to a broad, strong Fe K$\alpha$ line and negligible intrinsic absorption
the Fe L line interpretation could be plausible. Such a feature would
greatly constrain the IRR model parameters. 

We chose AX J1737.4-2907 as among the top four mainly because it is the brightest hard X-ray source in our sample, almost an order of magnitude brighter then the next brightest, LEDA 168563. Therefore, in spite of some possible complications due to a moderate absorption component, this source can be easily observed with any current hard X-ray missions. Moreover, it potentially has a very large Compton hump and a very broad Fe line feature.

\bigskip \noindent {\bf 5.2b Promising sources for Future Observations}

There are two more sources which satisfy the criteria for a promising source as described above
except that their hard X-ray flux level is rather low - IGR J16119-6036 and Mrk 1501.
These sources will be promising for the next generation X-ray observatories such as {\it ASTRO-H} with
higher hard X-ray sensitivity than, e.g., {\it Suzaku}.

IGR J16119-6036 has a distinct soft excess, a broader Fe line with
large EW, a low intrinsic absorption and a potential Compton hump.
The hard X-ray flux is somewhat low, but it would be well suited for the next generation
missions with better hard X-ray sensitivity such as {\it ASTHO-H}.

Mrk 1501 has a strong soft excess, a broader Fe line and the
absorption is very low. Its hard X-ray flux is somewhat low, but
due to the promising IRR features at lower energy bands it would be a good
candidate for the next generation X-ray missions with better hard X-ray sensitivity.

\subsection{\label{sub:other sources}Other Sources of Possible Interest}

Some additional sources were also identified as being of possible
interest. They were included because each of them has indication of 
at least one or more of some potentially interesting IRR features.
However, these sources are either more seriously contaminated by absorption
components or the possible features are less clear because of larger error bars 
in the critical regions due to the nature of short observations.

For IRAS 09149-6206 a simple model is unsuitable due to highly complicated
absorption features which seriously complicate matters. Also hard X-ray power 
law has a slope that does not match that of the softer band below 10.0 keV.
Nevertheless we found this source to be one of the more interesting ones 
with various reasons. First of all it appears to possess a board Fe line with 
a distinct asymmetric shape and a red wing which extends below 4.0 keV.
Also it has an indication of a distinct soft excess below 0.5 keV.
If fit only between 2.5 to 10.0 keV, a broad line extends from
7.0 keV down to 3.0 keV. This could be simply the effects
of the warm absorber extending above 2.0, which is plausible as seen
in Figure \ref{fig:ratio plots}. However if we assume that this
is not simply an effect of a warm absorber but instead the effect of
an extremely broadened Fe line, we can fit a broad Laor profile line
\citep{Laor:1991lr} with a narrow core superimposed on top. The result
is a statistically consistent model.

IRAS 05218+1212 and NVSS 193013+341047 both require large absorption components,
though they do not appear to be as complex as IRAS 09149-6206.
More importantly they appear to possess a board asymmetric Fe line with board red wings.
Moreover both these sources appear to possess large
Compton humps as the \textit{Swift}-BAT data is much stepper than the \textit{XMM-Newton} data.
Their apparent $\Gamma$ for \textit{Swift}-BAT data is closer to 2 \citep{Baumgartner:2013qy}, while for the \textit{XMM-Newton} data it is less than 1, see Table 2.

IGR J07597-3842 features a strong soft excess and possibly exhibits a strong Compton hump.  The source  has a low intrinsic absorption with a highly ionized absorber with a moderate column density.  These may make IGR J07597-3842 a potentially good source for \textit{ASTRO-H}. However it is hard to 
interpret the medium energy region due to large error bars.

NGC 4138, similar to IRAS 05218+1212 and NVSS 193013+341047,
likely has a large Compton hump. However this source exhibits the very flat spectrum 
in the \textit{XMM-Newton} band than others. More seriously, it has a very weak soft excess, 
if present at all, that is mostly absorbed by a large intrinsic absorption and moderately ionized column density. There is possibly a broad Fe line but large ionized absorber complicates the matter.

Fairall 1146 is well fit by IRR model with possible soft excess and hard Compton hump. The ionization of the IRR model is unconstrained and very low.  A large drop near 0.7 keV possibly indicates a strong warm absorber.   The model fit indicates a large
ionized absorption which complicates the matter when a longer observation or  higher resolution spectrum is obtained. A possible broad Fe K$\alpha$ line is hinted from the model fit,
but the large error bars near the region makes it unclear.

UGC 12138 is of possible interest due to the strong soft excess and possible Fe L line near 1.0 keV.
This could be either an O VII edge due to a warm absorber or a Fe L line.  The hard X-ray spectrum is flat, but soft excess is stronger and the EW of the Fe $\alpha$ K line is larger. However, given the short nature of the \textit{XMM-Newton} observation and large error bars the situation is
not clear.

\section{\label{sec:conclusion}Conclusion}

Using the latest \textit{Swift}-BAT data set \citep{Baumgartner:2013qy} we have
identified some lesser known AGN with short \textit{XMM-Newton} observations,
which we found most likely to exhibit some interesting prominent IRR features. 
All of these sources seem to be consistent with the IRR model.
A simple IRR model seems to fit most of the data set.

Of these sources we have identified four most prominent sources that
warrant farther scientific investigation by current missions: LEDA 168563, IGR 16558-5203,
SBS 1301+540 and AX J1737.4-2907. Should these sources prove to be
worthwhile, there are at least two other sources, IGR J16119-6036 and Mrk 1501, 
and potentially more, which could be good targets for the investigation
by the next generation X-ray missions, such as \textit{ASTRO-H}.

Due to only very short observation time of less than 20 ksec by \textit{XMM-Newton} available so far,
our general results, although encouraging, are yet to be confirmed. For instance, the black hole spin parameter could be constrained only for four of the sources in our list.
However, the prospect is very promising if longer observations are carried out, by currently running observatories such as \textit{Suzaku}, \textit{XMM-Newton}, and \textit{NuStar}, for our top four choices given in Section 5.2a, and by future observatories such as \textit{ASTRO-H} for our second choice targets given in Section 5.2b. Therefore, {\it these observations will be crucial} for our better understanding of the nature of the AGN central engine closest to the central black hole and the strong gravity.

\acknowledgments \textit{Acknowledgments}. This work is based on observations obtained with
\textit{XMM-Newton}, an ESA science mission with instruments and contributions
directly funded by ESA Member States and the USA (NASA).  This work has made use of NASA's Astrophysics Data System Bibliographic Services, HEASARC online databases, as well as of the NASA/IPAC Extragalactic Database (NED), which is operated by the Jet Propulsion Laboratory, California Institute of Technology, under contract with the National Aeronautics and Space Administration.

\textit{Facilities}:  Swift, XMM-Newton

\bibliographystyle{/Users/acl/Desktop/astronat/apj/apj}
\bibliography{hard_select_bib}

\begin{thebibliography}{34}
\expandafter\ifx\csname natexlab\endcsname\relax\def\natexlab#1{#1}\fi

\bibitem[{{Arnaud}(1996)}]{Arnaud:1996ud}
{Arnaud}, K.~A. 1996, in Astronomical Society of the Pacific Conference Series,
  Vol. 101, Astronomical Data Analysis Software and Systems V, ed.
  {G.~H.~Jacoby \& J.~Barnes}, 17--+

\bibitem[{{Baumgartner} {et~al.}(2013){Baumgartner}, {Tueller}, {Markwardt},
  {Skinner}, {Barthelmy}, {Mushotzky}, {Evans}, \&
  {Gehrels}}]{Baumgartner:2013qy}
{Baumgartner}, W.~H., {Tueller}, J., {Markwardt}, C.~B., {Skinner}, G.~K.,
  {Barthelmy}, S., {Mushotzky}, R.~F., {Evans}, P.~A., \& {Gehrels}, N. 2013,
  ApJS, 207, 19

\bibitem[{{Crummy} {et~al.}(2006){Crummy}, {Fabian}, {Gallo}, \&
  {Ross}}]{Crummy:2006ly}
{Crummy}, J., {Fabian}, A.~C., {Gallo}, L., \& {Ross}, R.~R. 2006, \mnras, 365,
  1067

\bibitem[{{den Herder} {et~al.}(2001)}]{den-Herder:2001nx}
{den Herder}, J.~W. {et~al.} 2001, A\&A, 365, L7

\bibitem[{{Done} {et~al.}(2012){Done}, {Davis}, {Jin}, {Blaes}, \&
  {Ward}}]{Done:2012cr}
{Done}, C., {Davis}, S.~W., {Jin}, C., {Blaes}, O., \& {Ward}, M. 2012, \mnras,
  420, 1848

\bibitem[{{Fabian} {et~al.}(1989){Fabian}, {Rees}, {Stella}, \&
  {White}}]{Fabian:1989uq}
{Fabian}, A.~C., {Rees}, M.~J., {Stella}, L., \& {White}, N.~E. 1989, \mnras,
  238, 729

\bibitem[{{Fabian} {et~al.}(2009){Fabian}, {Zoghbi}, {Ross}, {Uttley}, {Gallo},
  {Brandt}, {Blustin}, {Boller}, {Caballero-Garcia}, {Larsson}, {Miller},
  {Miniutti}, {Ponti}, {Reis}, {Reynolds}, {Tanaka}, \&
  {Young}}]{Fabian:2009fk}
{Fabian}, A.~C., {Zoghbi}, A., {Ross}, R.~R., {Uttley}, P., {Gallo}, L.~C.,
  {Brandt}, W.~N., {Blustin}, A.~J., {Boller}, T., {Caballero-Garcia}, M.~D.,
  {Larsson}, J., {Miller}, J.~M., {Miniutti}, G., {Ponti}, G., {Reis}, R.~C.,
  {Reynolds}, C.~S., {Tanaka}, Y., \& {Young}, A.~J. 2009, \nat, 459, 540

\bibitem[{{Gierli{\'n}ski} \& {Done}(2004)}]{Gierlinski:2004dq}
{Gierli{\'n}ski}, M. \& {Done}, C. 2004, \mnras, 349, L7

\bibitem[{{Jin} {et~al.}(2012){Jin}, {Ward}, {Done}, \& {Gelbord}}]{Jin:2012ys}
{Jin}, C., {Ward}, M., {Done}, C., \& {Gelbord}, J. 2012, \mnras, 420, 1825

\bibitem[{{Kalberla} {et~al.}(2005){Kalberla}, {Burton}, {Hartmann}, {Arnal},
  {Bajaja}, {Morras}, \& {P{\"o}ppel}}]{Kalberla:2005kx}
{Kalberla}, P.~M.~W., {Burton}, W.~B., {Hartmann}, D., {Arnal}, E.~M.,
  {Bajaja}, E., {Morras}, R., \& {P{\"o}ppel}, W.~G.~L. 2005, A\&A, 440, 775

\bibitem[{{Kara} {et~al.}(2013{\natexlab{a}}){Kara}, {Fabian}, {Cackett},
  {Steiner}, {Uttley}, {Wilkins}, \& {Zoghbi}}]{Kara:2013ys}
{Kara}, E., {Fabian}, A.~C., {Cackett}, E.~M., {Steiner}, J.~F., {Uttley}, P.,
  {Wilkins}, D.~R., \& {Zoghbi}, A. 2013{\natexlab{a}}, \mnras, 428, 2795

\bibitem[{{Kara} {et~al.}(2013{\natexlab{b}}){Kara}, {Fabian}, {Cackett},
  {Uttley}, {Wilkins}, \& {Zoghbi}}]{Kara:2013vn}
{Kara}, E., {Fabian}, A.~C., {Cackett}, E.~M., {Uttley}, P., {Wilkins}, D.~R.,
  \& {Zoghbi}, A. 2013{\natexlab{b}}, \mnras, 434, 1129

\bibitem[{{Laor}(1991)}]{Laor:1991lr}
{Laor}, A. 1991, ApJ, 376, 90

\bibitem[{{Magdziarz} \& {Zdziarski}(1995)}]{Magdziarz:1995dp}
{Magdziarz}, P. \& {Zdziarski}, A.~A. 1995, MNRAS, 273, 837

\bibitem[{{Middleton} {et~al.}(2007){Middleton}, {Done}, \&
  {Gierli{\'n}ski}}]{Middleton:2007vn}
{Middleton}, M., {Done}, C., \& {Gierli{\'n}ski}, M. 2007, \mnras, 381, 1426

\bibitem[{{Miller} {et~al.}(2008){Miller}, {Turner}, \&
  {Reeves}}]{Miller:2008ve}
{Miller}, L., {Turner}, T.~J., \& {Reeves}, J.~N. 2008, \aap, 483, 437

\bibitem[{{Miller} {et~al.}(2009){Miller}, {Turner}, \&
  {Reeves}}]{Miller:2009qf}
---. 2009, \mnras, 399, L69

\bibitem[{{Miller} {et~al.}(2010){Miller}, {Turner}, {Reeves}, \&
  {Braito}}]{Miller:2010zr}
{Miller}, L., {Turner}, T.~J., {Reeves}, J.~N., \& {Braito}, V. 2010, \mnras,
  408, 1928

\bibitem[{{Miniutti} \& {Fabian}(2004)}]{Miniutti:2004lr}
{Miniutti}, G. \& {Fabian}, A.~C. 2004, MNRAS, 349, 1435

\bibitem[{{Miniutti} {et~al.}(2007){Miniutti}, {Fabian}, {Anabuki}, {Crummy},
  {Fukazawa}, {Gallo}, {Haba}, {Hayashida}, {Holt}, {Kunieda}, {Larsson},
  {Markowitz}, {Matsumoto}, {Ohno}, {Reeves}, {Takahashi}, {Tanaka},
  {Terashima}, {Torii}, {Ueda}, {Ushio}, {Watanabe}, {Yamauchi}, \&
  {Yaqoob}}]{Miniutti:2007bh}
{Miniutti}, G., {Fabian}, A.~C., {Anabuki}, N., {Crummy}, J., {Fukazawa}, Y.,
  {Gallo}, L., {Haba}, Y., {Hayashida}, K., {Holt}, S., {Kunieda}, H.,
  {Larsson}, J., {Markowitz}, A., {Matsumoto}, C., {Ohno}, M., {Reeves}, J.~N.,
  {Takahashi}, T., {Tanaka}, Y., {Terashima}, Y., {Torii}, K., {Ueda}, Y.,
  {Ushio}, M., {Watanabe}, S., {Yamauchi}, M., \& {Yaqoob}, T. 2007, \pasj, 59,
  315

\bibitem[{{Nandra} {et~al.}(1999){Nandra}, {George}, {Mushotzky}, {Turner}, \&
  {Yaqoob}}]{Nandra:1999kx}
{Nandra}, K., {George}, I.~M., {Mushotzky}, R.~F., {Turner}, T.~J., \&
  {Yaqoob}, T. 1999, \apjl, 523, L17

\bibitem[{{Nandra} {et~al.}(2007){Nandra}, {O'Neill}, {George}, \&
  {Reeves}}]{Nandra:2007jk}
{Nandra}, K., {O'Neill}, P.~M., {George}, I.~M., \& {Reeves}, J.~N. 2007,
  MNRAS, 382, 194

\bibitem[{{Ricci} {et~al.}(2011){Ricci}, {Walter}, {Courvoisier}, \&
  {Paltani}}]{Ricci:2011ly}
{Ricci}, C., {Walter}, R., {Courvoisier}, T.~J.-L., \& {Paltani}, S. 2011,
  \aap, 532, A102

\bibitem[{{Ross} \& {Fabian}(2005)}]{Ross:2005qe}
{Ross}, R.~R. \& {Fabian}, A.~C. 2005, MNRAS, 358, 211

\bibitem[{{Str{\"u}der} {et~al.}(2001)}]{Struder:2001cr}
{Str{\"u}der}, L. {et~al.} 2001, A\&A, 365, L18

\bibitem[{{Tanaka} {et~al.}(1995){Tanaka}, {Nandra}, {Fabian}, {Inoue},
  {Otani}, {Dotani}, {Hayashida}, {Iwasawa}, {Kii}, {Kunieda}, {Makino}, \&
  {Matsuoka}}]{Tanaka:1995oq}
{Tanaka}, Y., {Nandra}, K., {Fabian}, A.~C., {Inoue}, H., {Otani}, C.,
  {Dotani}, T., {Hayashida}, K., {Iwasawa}, K., {Kii}, T., {Kunieda}, H.,
  {Makino}, F., \& {Matsuoka}, M. 1995, Nature, 375, 659

\bibitem[{{Tueller} {et~al.}(2008){Tueller}, {Mushotzky}, {Barthelmy},
  {Cannizzo}, {Gehrels}, {Markwardt}, {Skinner}, \& {Winter}}]{Tueller:2008qf}
{Tueller}, J., {Mushotzky}, R.~F., {Barthelmy}, S., {Cannizzo}, J.~K.,
  {Gehrels}, N., {Markwardt}, C.~B., {Skinner}, G.~K., \& {Winter}, L.~M. 2008,
  \apj, 681, 113

\bibitem[{{Turner} {et~al.}(2001)}]{Turner:2001rr}
{Turner}, M.~J.~L. {et~al.} 2001, A\&A, 365, L27

\bibitem[{{Walton} {et~al.}(2013){Walton}, {Nardini}, {Fabian}, {Gallo}, \&
  {Reis}}]{Walton:2013qy}
{Walton}, D.~J., {Nardini}, E., {Fabian}, A.~C., {Gallo}, L.~C., \& {Reis},
  R.~C. 2013, MNRAS, 428, 2901

\bibitem[{{Walton} {et~al.}(2012){Walton}, {Reis}, {Cackett}, {Fabian}, \&
  {Miller}}]{Walton:2012fk}
{Walton}, D.~J., {Reis}, R.~C., {Cackett}, E.~M., {Fabian}, A.~C., \& {Miller},
  J.~M. 2012, \mnras, 422, 2510

\bibitem[{{Winter} {et~al.}(2010){Winter}, {Lewis}, {Koss}, {Veilleux},
  {Keeney}, \& {Mushotzky}}]{Winter:2010ve}
{Winter}, L.~M., {Lewis}, K.~T., {Koss}, M., {Veilleux}, S., {Keeney}, B., \&
  {Mushotzky}, R.~F. 2010, \apj, 710, 503

\bibitem[{{Winter} {et~al.}(2009){Winter}, {Mushotzky}, {Reynolds}, \&
  {Tueller}}]{Winter:2009zr}
{Winter}, L.~M., {Mushotzky}, R.~F., {Reynolds}, C.~S., \& {Tueller}, J. 2009,
  \apj, 690, 1322

\bibitem[{{Zoghbi} \& {Fabian}(2011)}]{Zoghbi:2011uq}
{Zoghbi}, A. \& {Fabian}, A.~C. 2011, \mnras, 418, 2642

\bibitem[{{Zoghbi} {et~al.}(2012){Zoghbi}, {Fabian}, {Reynolds}, \&
  {Cackett}}]{Zoghbi:2012kx}
{Zoghbi}, A., {Fabian}, A.~C., {Reynolds}, C.~S., \& {Cackett}, E.~M. 2012,
  \mnras, 422, 129

\end{thebibliography}

%

%
\begin{table}
\caption{\label{tab:obs log}Observation log of hard X-ray selected AGN and
their basic properties.}

\scalebox{0.90}{
\begin{tabular}{ccccccc}
\hline
Source  & RA  & Dec  & $z$  & Observation ID  & Observation Date  & Exposure\tabularnewline
 & (J2000.0)  & (J2000.0)  &  &  &  & (s)\tabularnewline
\hline
LEDA 168563  & $04^{\text{h}}52^{\text{m}}05.00^{\text{s}}$  & $+49$\textdegree{} $ 32' 45.0''$  & 0.0290  & 0401790201  & 2007 February 26  & 12413\tabularnewline
IGR J16558-5203  & 16$^{\text{h}}$56$^{\text{m}}$05.60$^{\text{s}}$  & $-52$\textdegree{} $03' 41.0''$  & 0.0540  & 0306171201  & 2006 March 1  & 11817\tabularnewline
SBS 1301+540  & 13$^{\text{h}}$03$^{\text{m}}$59.47$^{\text{s}}$  & $+53$\textdegree{} $47' 30.1''$  & 0.0299  & 0312192001  & 2006 June 23  & 11912\tabularnewline
AX J1737.4-2907  & 17$^{\text{h}}$37$^{\text{m}}$28.31$^{\text{s}}$  & $-29$\textdegree{} $08' 02.4''$  & 0.0214  & 0550451501  & 2009 February 26  & 17919\tabularnewline
IGR J16119-6036  & 16$^{\text{h}}$11$^{\text{m}}$51.40$^{\text{s}}$  & $-60$\textdegree{} $37' 55.0''$  & 0.0156  & 0550451101  & 2009 February 18  & 17918\tabularnewline
Mrk 1501  & 00$^{\text{h}}$10$^{\text{m}}$31.00$^{\text{s}}$  & $+10$\textdegree{} $58' 30.0''$  & 0.0893  & 0127110201  & 2000 July 03  & 16311\tabularnewline
IRAS 09149-6206  & 09$^{\text{h}}$16$^{\text{m}}$09.40$^{\text{s}}$  & $-62$\textdegree{} $19' 30.0'' $ & 0.0573  & 0550452601  & 2008 December 14  & 17815\tabularnewline
IRAS 05218-1212 & 05$^{\text{h}}$24$^{\text{m}}$06.40$^{\text{s}}$ & $-12$\textdegree{} $09' 27.0''$ & 0.0490 & 0551950401 & 2008 August 25 & 14918\tabularnewline
NVSS 193013+341047 & 19$^{\text{h}}$30$^{\text{m}}$13.85$^{\text{s}}$ & $+34$\textdegree{} $10' 50.0''$ & 0.0629 & 0602840101 & 2009 May 16 & 16916\tabularnewline
IGR J07597-3842  & 07$^{\text{h}}$59$^{\text{m}}$41.60$^{\text{s}}$  & $-38$\textdegree{} $43' 57.0''$  & 0.0400  & 0303230101  & 2006 April 8  & 16815\tabularnewline
NGC 4183 & 12$^{\text{h}}$09$^{\text{m}}$29.79$^{\text{s}}$ & $+43$\textdegree{} $41' 07.1''$ & 0.0030 & 0112551201 & 2001 November 26 & 14965\tabularnewline
Fairall 1146 & 08$^{\text{h}}$38$^{\text{m}}$30.79$^{\text{s}}$ & $-35$\textdegree{} $59' 33.7''$ & 0.0316 & 0401790401 & 2006 December 12 & 17420\tabularnewline
UGC 12138 & 22$^{\text{h}}$40$^{\text{m}}$17.00$^{\text{s}}$ & $+08$\textdegree{} $03' 14.0''$ & 0.0250 & 0103860301 & 2001 June 3 & 8105\tabularnewline
\hline
\end{tabular}
}
\end{table}

%

%
\begin{table}[H]
\caption{\label{tab:simple pl}Simple power law fit parameters}

\scalebox{0.90}{
\begin{tabular}{ccccc}
\hline
Source  & $\Gamma$  & $N_{H}$  & Flux (2.0-10.0 keV)  & $\chi^{2}$(d.o.f.)\tabularnewline
 &  & ($10^{22}$ cm$^{-2}$)  & ($10^{-12}$ erg cm$^{-2}$ s$^{-1}$)  & \tabularnewline
\hline
LEDA 168563  & $1.75_{-0.02}^{+0.02}$  & $0.542$  & $4.06_{-0.03}^{+0.03}$  & $251.41\:(205)$\tabularnewline
IGR J16558-5203  & $1.80_{-0.05}^{+0.05}$  & $0.246$  & $0.99_{-0.01}^{+0.01}$  & $468.68\:470$\tabularnewline
SBS 1301+540  & $1.69_{-0.03}^{+0.04}$  & $1.69e-2$  & $1.50_{-0.01}^{+0.02}$  & $184.51\:(170)$\tabularnewline
AX J1737.4-2907  & $1.28_{-0.02}^{+0.02}$  & $0.757$  & $5.18_{-0.04}^{+0.05}$  & $273.12\:(210)$\tabularnewline
IGR J16119-6036  & $1.62_{-0.06}^{+0.06}$  & $0.183$  & $0.53_{-0.01}^{+0.01}$  & $354.57\:(339)$\tabularnewline
Mrk 1501  & $1.60_{-0.06}^{+0.06}$  & $5.39e-2$  & $0.65_{-0.01}^{+0.01}$  & $265.31\:(262)$\tabularnewline
IRAS 09149-6206  & $1.16_{-0.03}^{+0.03}$  & $0.158$  & $0.85_{-0.01}^{+0.01}$  & $973.36\:(613)$\tabularnewline
IRAS 05218-1212 & $0.36_{-0.16}^{+0.16}$ & $9.75e-2$ & $2.39_{-0.15}^{+0.17}$ & $44.41\:(36)$\tabularnewline
NVSS 193013+341047 & $0.56_{-0.16}^{+0.15}$ & $0.158$ & $2.61_{-0.14}^{+0.10}$ & $17.66\:(26)$\tabularnewline
IGR J07597-3842  & $1.53_{-0.03}^{+0.03}$  & $0.536$  & $1.53_{-0.02}^{+0.02}$  & $695.81\:(702)$\tabularnewline
NGC 4183 & $0.72_{-0.10}^{+0.10}$ & $1.25e-2$ & $5.70_{-0.13}^{+0.10}$ & $196.83\:(124)$\tabularnewline
Fairall 1146 & $1.59_{-0.01}^{+0.01}$ & $0.365$ & $12.31_{-0.17}^{+0.16}$ & $203.67\:(214)$\tabularnewline
UGC 12138 & $1.82_{-0.13}^{+0.13}$ & $6.39e-2$ & $7.40_{-0.32}^{+0.28}$ & $59.57\:(69)$\tabularnewline
\hline
\end{tabular}
}
\end{table}

%

%
\begin{table}
\caption{\label{tab:spectal fits}Results of spectral fitting}

\scalebox{0.85}{
\begin{tabular}{cccccccc}
\hline
Source  & $\Gamma$  & $N_{H}^{int}$  & $\xi$  & q  & a  & R  & $\chi^{2}$(d.o.f.)\tabularnewline
 &  & ($10^{22}$ cm$^{-2}$)  & (erg cm s$^{-1}$)  &  &  &  & \tabularnewline
\hline
LEDA 168563  & $1.98_{-0.01}^{+0.01}$  & $1.26_{-0.08}^{+0.08}\times10^{-1}$  & $203.0_{-11.3}^{+4.0}$  & $4.39_{-0.50}^{+0.53}$  & $0.62_{-0.02}^{+0.07}$  & $-0.27_{-0.13}^{+0.13}$  & $2250.40\:(2107)$\tabularnewline
IGR J16558-5203  & $2.05_{-0.02}^{+0.02}$  & $9.90_{-1.37}^{+1.36}\times10^{-2}$  & $235.1_{-18.6}^{+28.8}$  & $7.04_{-1.69}^{+2.58}$  & $0.71_{-0.05}^{+0.06}$  & $-1.08_{-0.30}^{+0.30}$  & $1149.65\:(1127)$\tabularnewline
SBS 1301+540  & $1.83_{-0.01}^{+0.01}$  & -  & $40.9_{-16.9}^{+10.6}$  & $6.01_{-0.72}^{+0.74}$  & $<0.99$  & $-0.38_{-0.16}^{+0.16}$  & $1483.93\:(1476)$\tabularnewline
AX J1737.4-2907  & $1.89_{-0.02}^{+0.02}$  & $8.83_{-0.20}^{+0.20}\times10^{-1}$  & $109.8_{-43.6}^{+59.8}$  & $7.51_{-0.59}^{+0.72}$  & $0.97_{-0.02}^{+0.01}$  & $-0.32_{-0.16}^{+0.16}$  & $1888.55\:(1895)$\tabularnewline
IGR J16119-6036  & $1.84_{-0.02}^{+0.02}$  & $4.72_{-1.04}^{+1.01}\times10^{-2}$  & $70.2_{-9.8}^{+25.6}$  & $4.36_{-0.51}^{+0.49}$  & $<0.99$  & $-1.07_{-0.33}^{+0.33}$  & $888.84\:(903)$\tabularnewline
Mrk 1501  & $1.66_{-0.03}^{+0.03}$  & $3.38_{-0.77}^{+0.76}\times10^{-2}$  & $690.3_{-172.9}^{+362.6}$  & $6.05_{-1.27}^{+1.85}$  & $0.85_{-0.05}^{+0.04}$  & $<-0.35$  & $1701.99\:(1703)$\tabularnewline
IRAS 09149-6206   & $1.68_{-0.03}^{+0.03}$  & -  & $498.0_{-245.4}^{+87.0}$  & $<8.11$  & $<0.50$  & $-0.63_{-0.19}^{+0.20}$  & $1273.68\:(1224)$\tabularnewline
IRAS 05218-1212 & $1.81_{-0.04}^{+0.04}$ & $6.57_{-0.01}^{+0.02}\times10^{-2}$ & $49.3_{-38.4}^{+10.4}$ & $<9.66$ & $<0.79$ & $-1.16_{-0.92}^{+0.87}$ & $311.11\:(269)$\tabularnewline
NVSS 193013+341047 & $1.64_{-0.03}^{+0.04}$ & $3.38_{-0.38}^{+0.39}\times10^{-1}$ & $119.2_{-14.1}^{+27.6}$ & $<9.90$ & $<0.98$ & $-0.57_{-0.59}^{+0.57}$ & $238.71\:(202)$\tabularnewline
IGR J07597-3842  & $1.73_{-0.01}^{+0.01}$  & $9.38_{-0.92}^{+0.92}\times10^{-2}$  & $322.4_{-32.4}^{+50.9}$  & $7.61_{-0.46}^{+0.45}$  & $<0.98$  & $-0.86_{-0.19}^{+0.19}$  & $1672.97\:(1682)$\tabularnewline
NGC 4183 & $1.87_{-0.04}^{+0.04}$ & $2.49_{-0.23}^{+0.26}\times10^{-1}$ & $198.3_{-28.8}^{+4.0}$ & $<9.90$ & $<0.98$ & $<-0.82$ & $504.33\:(358)$\tabularnewline
Fairall 1146 & $1.98_{-0.01}^{+0.01}$ & $1.38_{-0.01}^{+0.01}\times10^{-2}$ & $<15.8$ & $2.48_{-0.24}^{+0.26}$ & $<0.91$ & $-0.43_{-0.26}^{+0.26}$ & $450.20\:(388)$\tabularnewline
UGC 12138 & $2.06_{-0.05}^{+0.03}$ & $1.26_{-0.01}^{+0.01}\times10^{-2}$ & $58.9_{-8.4}^{+41.9}$ & $5.10_{-0.52}^{+0.60}$ & $<0.93$ & $-0.91_{-0.77}^{+0.75}$ & $724.10\:(733)$\tabularnewline
\hline
\end{tabular}
}
\end{table}

%

%
\begin{table}

\caption{\label{tab:mod parameteres}Modifications to simple model}

\scalebox{0.90}{
\begin{tabular}{cccccc}
\hline
Source  & $N_{H}^{int}$  & $N_{H}^{pc}$  & $log\:\xi^{pc}$  &  & $\Delta\chi^{2}$\tabularnewline
 & ($10^{22}$ cm$^{-2}$)  & ($10^{22}$ cm$^{-2}$)  &  &  & \tabularnewline
\hline
LEDA 168563  & $1.26_{-0.08}^{+0.08}\times10^{-1}$  & $3.63_{-1.35}^{+1.12}\times10^{-1}$  & $3.05_{-0.12}^{+0.12}$  &  & $29.80$\tabularnewline
IGR J16558-5203  & $9.90_{-1.37}^{+1.36}\times10^{-2}$  & $<7.96\times10^{-2}$  & $2.74_{-0.36}^{+0.58}$  &  & $3.50$\tabularnewline
SBS 1301+540  & -  & -  & -  &  & \tabularnewline
AX J1737.4-2907  & $8.83_{-0.20}^{+0.20}\times10^{-1}$  & $1.26_{-0.11}^{+0.12}$  & $2.17_{-0.07}^{+0.07}$  &  & $19.79$\tabularnewline
IGR J16119-6036  & $4.72_{-1.04}^{+1.01}\times10^{-2}$  & -  & -  &  & \tabularnewline
Mrk 1501  & $3.38_{-0.77}^{+0.76}\times10^{-2}$  & -  & -  &  & \tabularnewline
IRAS 09149-6206   & -  & $2.47_{-0.06}^{+0.06}$  & $0.33_{-0.01}^{+0.01}$  &  & \tabularnewline
IRAS 05218-1212 & $6.57_{-0.01}^{+0.02}\times10^{-2}$ & $17.52_{-0.80}^{+0.76}$ & $2.01_{-0.01}^{+0.01}$ &  & $699.23$\tabularnewline
NVSS 193013+341047 & $3.38_{-0.38}^{+0.39}\times10^{-1}$ & $46.15_{-2.67}^{+2.41}$ & $2.07_{-0.02}^{+0.02}$ &  & $32.55$\tabularnewline
IGR J07597-3842  & $9.38_{-0.92}^{+0.92}\times10^{-2}$  & $1.03_{-0.51}^{+0.53}$  & $3.22_{-0.12}^{+0.12}$  &  & $6.68$\tabularnewline
NGC 4183 & $2.49_{-0.23}^{+0.26}\times10^{-1}$ & $20.22_{-0.26}^{+0.25}$ & $1.92_{-0.02}^{+0.01}$ &  & $>1000$\tabularnewline
Fairall 1146 & $1.38_{-0.01}^{+0.01}\times10^{-2}$ & $1.20_{-0.04}^{+0.04}$ & $1.35_{-0.02}^{+0.02}$ &  & $700.90$\tabularnewline
UGC 12138 & $1.26_{-0.01}^{+0.01}\times10^{-2}$ & $1.58_{-0.39}^{+0.36}\times10^{-1}$ & $1.64_{-0.17}^{+0.18}$ &  & $26.43$\tabularnewline
\hline
\end{tabular}
}
\end{table}

%

%
\begin{table}
\caption{\label{tab:Fe line table}Fe line energies and equivalent widths.}

\scalebox{0.90}{
\begin{tabular}{ccc}
\hline
Source  & Line Energy  & EW \tabularnewline
 & (keV)  & (eV)\tabularnewline
\hline
LEDA 168563  & $6.45_{-0.06}^{+0.05}$  & $38.0_{-18.6}^{+18.6}$\tabularnewline
IGR J16558-5203  & $6.49_{-0.06}^{+0.07}$  & $115.1_{-53.0}^{+55.8}$\tabularnewline
SBS 1301+540  & $6.98_{-0.07}^{+0.06}$  & $78.5_{-42.2}^{+42.5}$\tabularnewline
AX J1737.4-2907  & $5.99_{-0.10}^{+0.14}$  & $29.2_{-21.1}^{+23.4}$\tabularnewline
IGR J16119-6036  & $6.48_{-0.07}^{+0.06}$  & $137.6_{-75.2}^{+74.8}$\tabularnewline
Mrk 1501  & $6.70_{-0.43}^{+0.37}$  & $102.3_{-74.1}^{+82.9}$\tabularnewline
IRAS 09149-6206  & $6.44_{-0.05}^{+0.05}$  & $138.2_{-37.7}^{+40.6}$\tabularnewline
IRAS 05218-1212 & $6.40_{-0.05}^{+0.03}$ & $348.2_{-147.1}^{+166.4}$\tabularnewline
NVSS 193013+341047 & $6.41_{-0.06}^{+0.05}$ & $109.3_{-70.9}^{+72.1}$\tabularnewline
IGR J07597-3842  & $6.44_{-0.05}^{+0.04}$  & $70.6_{-23.9}^{+27.0}$\tabularnewline
NGC 4183 & $6.35_{-0.05}^{+0.04}$ & $119.3_{-57.7}^{+62.6}$\tabularnewline
Fairall 1146 & $6.39_{-0.09}^{+0.09}$ & $170.6_{-75.3}^{+68.0}$\tabularnewline
UGC 12138 & $6.31_{-0.10}^{+0.10}$ & $132.5_{-93.1}^{+112.7}$\tabularnewline
\hline
\end{tabular}
}
\end{table}

\begin{figure}[h]
\includegraphics[bb=0bp 0bp 792bp 512bp,scale=0.3]{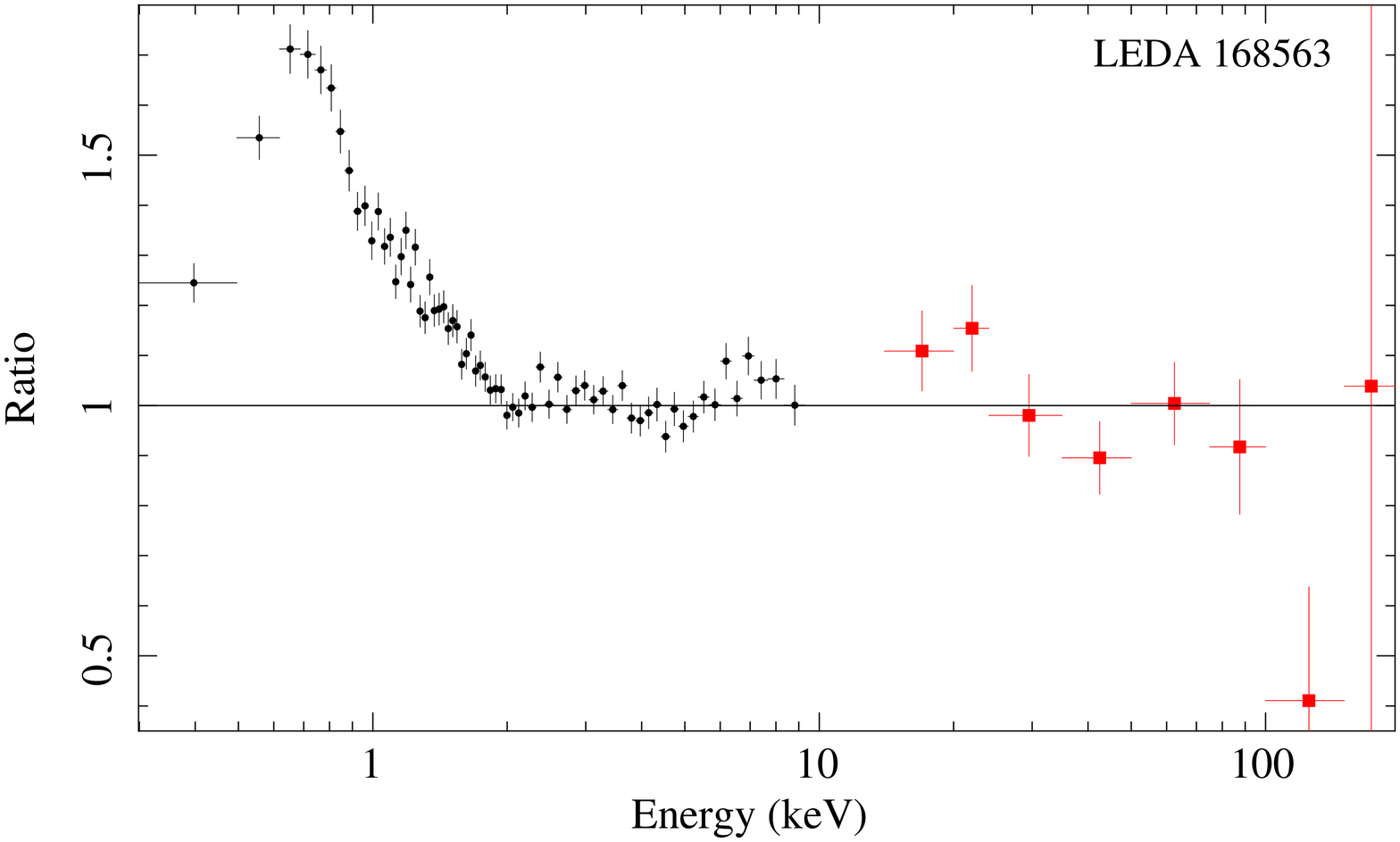}\includegraphics[bb=0bp 0bp 792bp 512bp,scale=0.3]{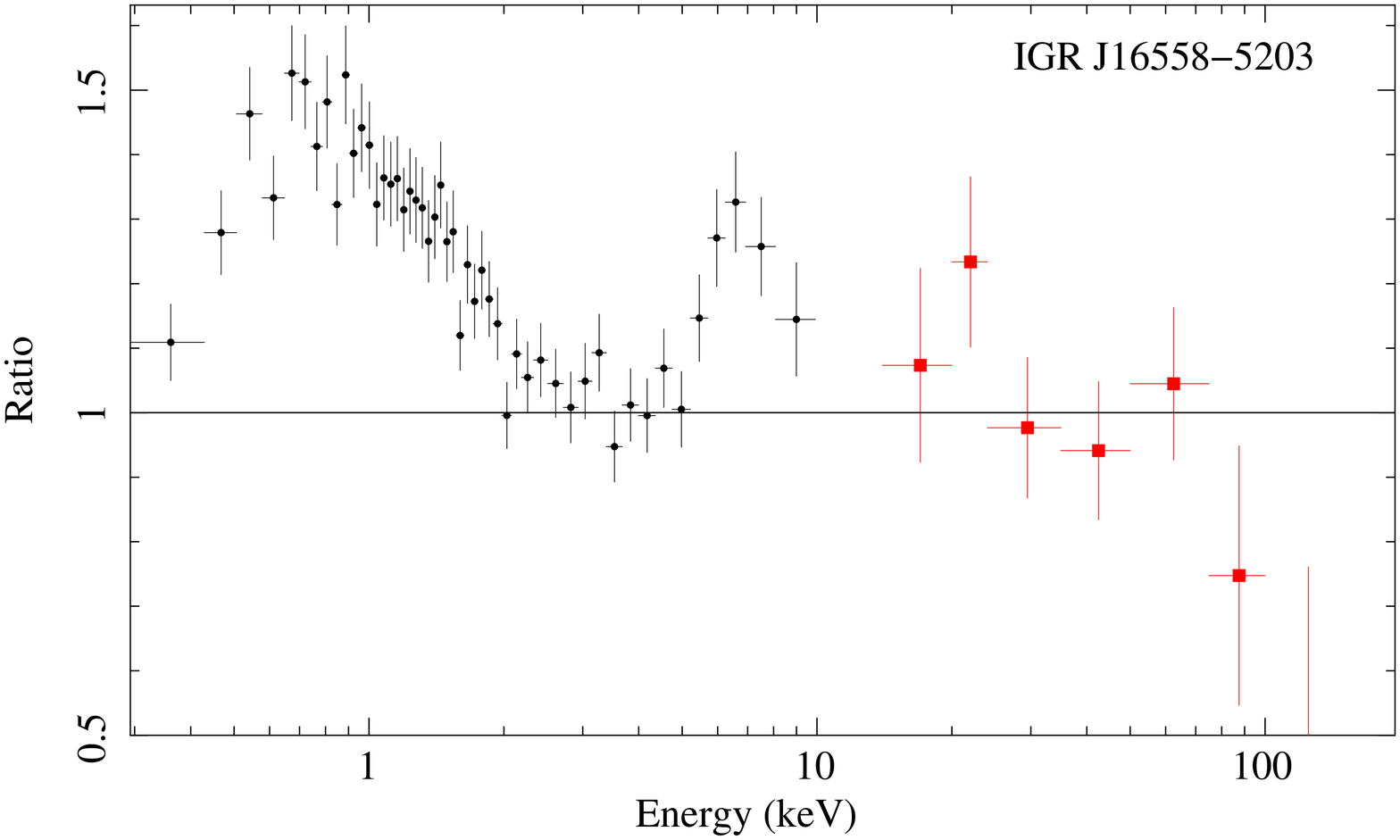}

\includegraphics[bb=0bp 0bp 792bp 512bp,scale=0.3]{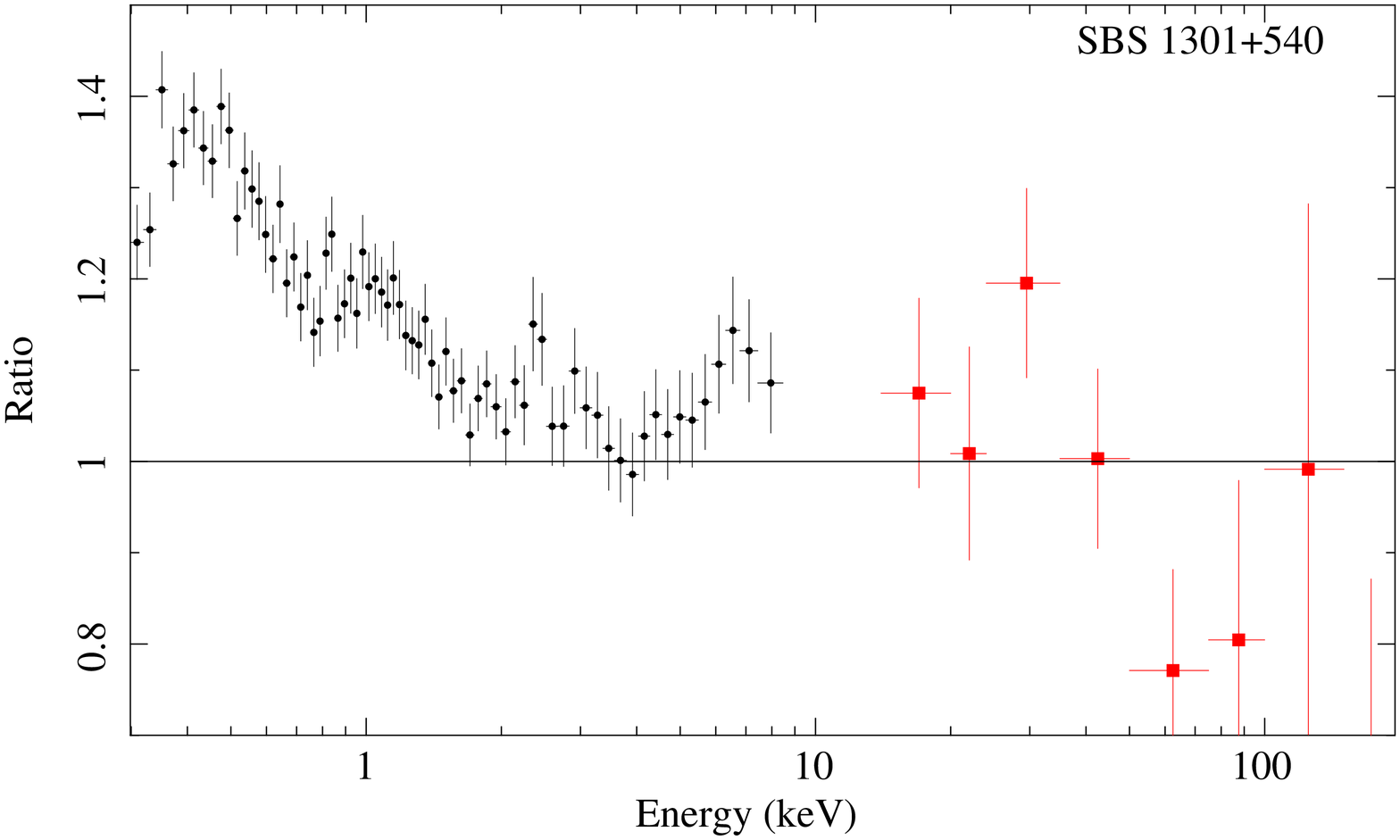}\includegraphics[bb=0bp 0bp 792bp 512bp,scale=0.3]{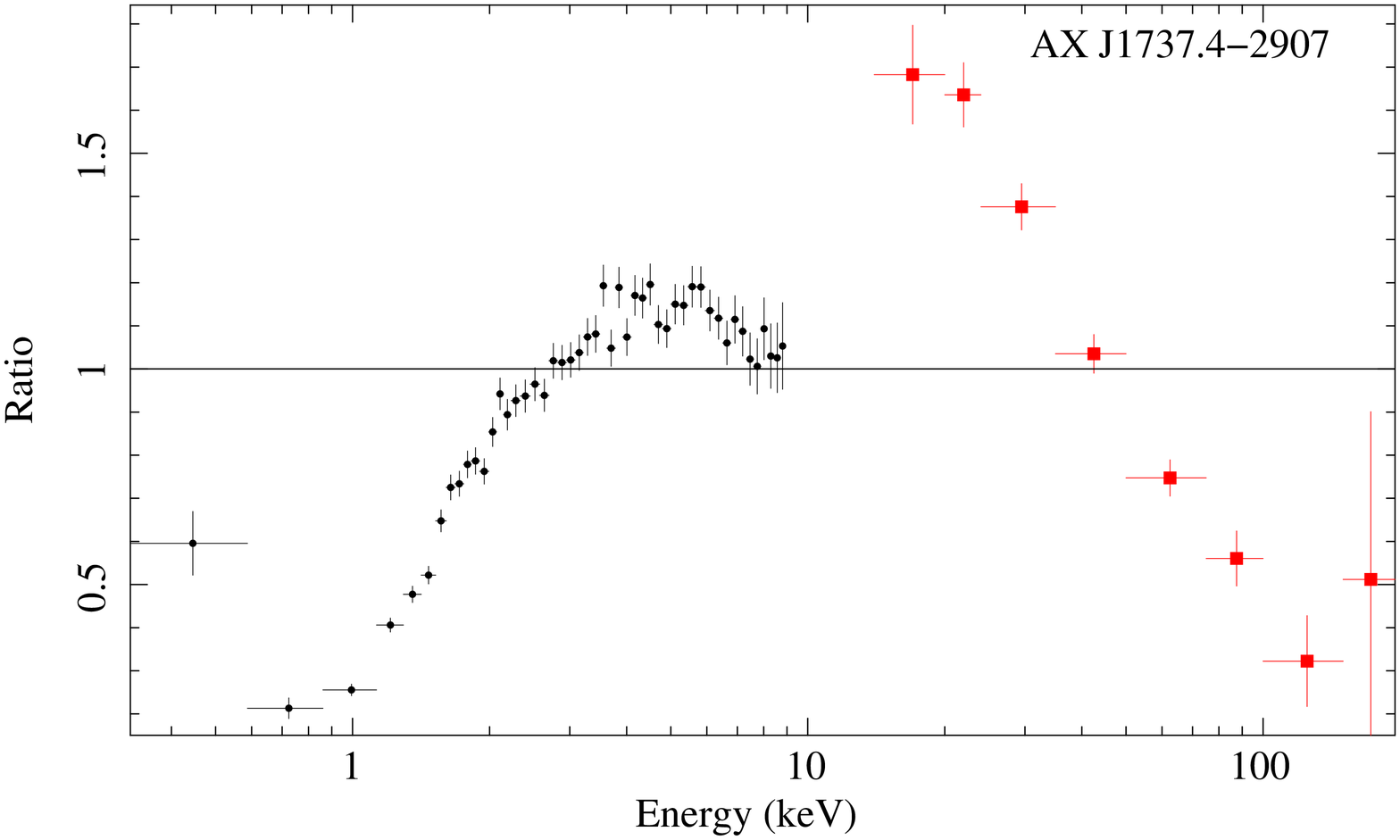}

\includegraphics[bb=0bp 0bp 792bp 512bp,scale=0.3]{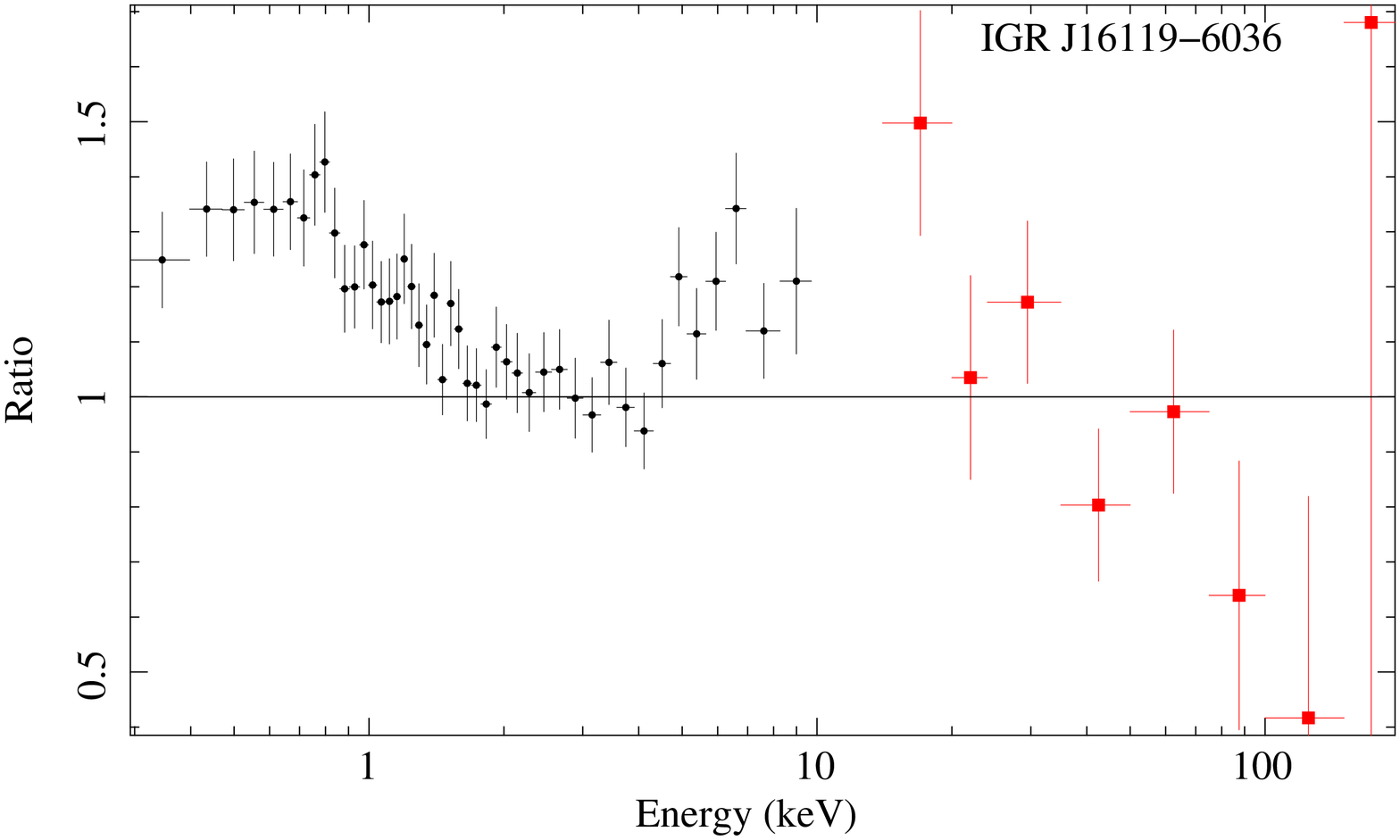}\includegraphics[bb=0bp 0bp 792bp 512bp,scale=0.3]{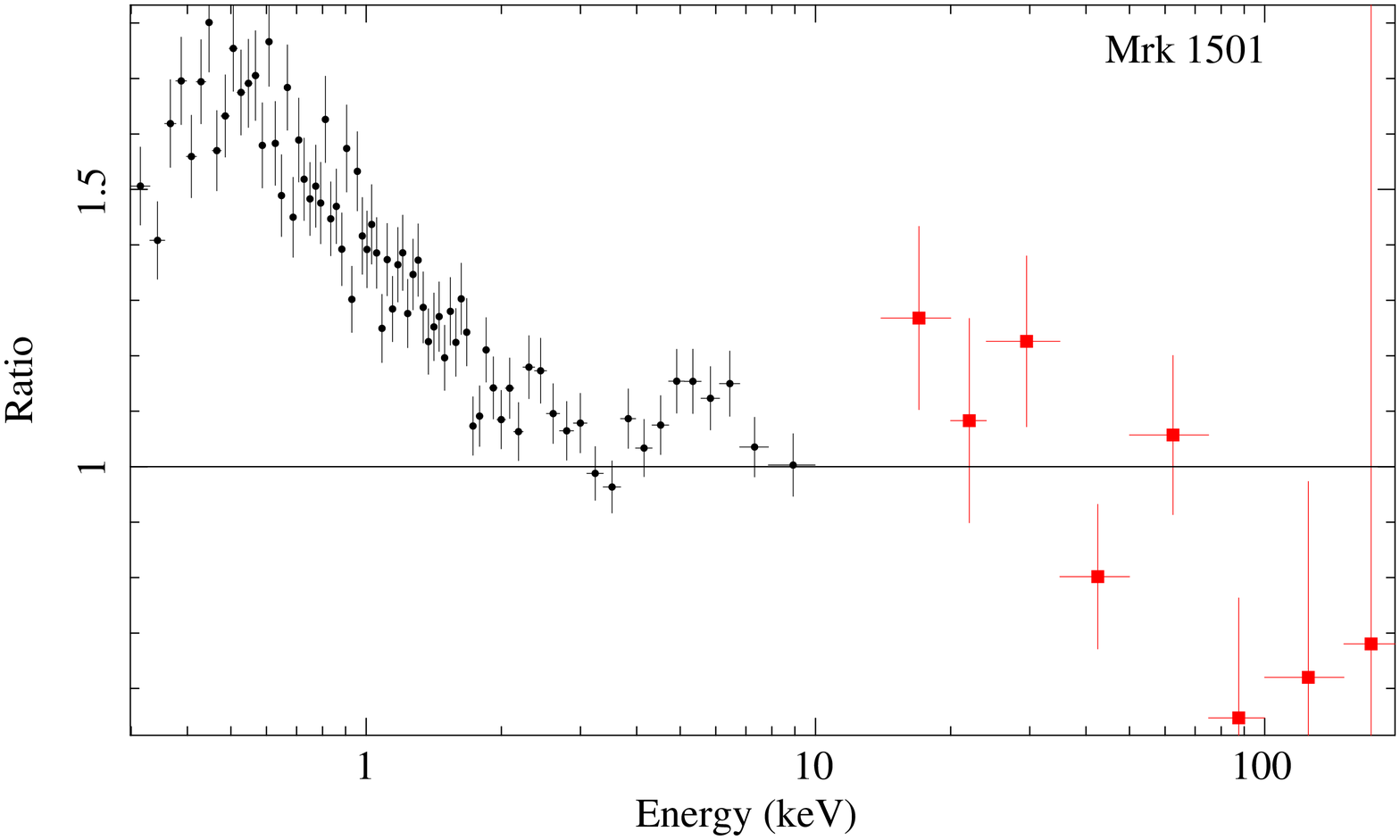}

\includegraphics[bb=0bp 0bp 792bp 512bp,scale=0.3]{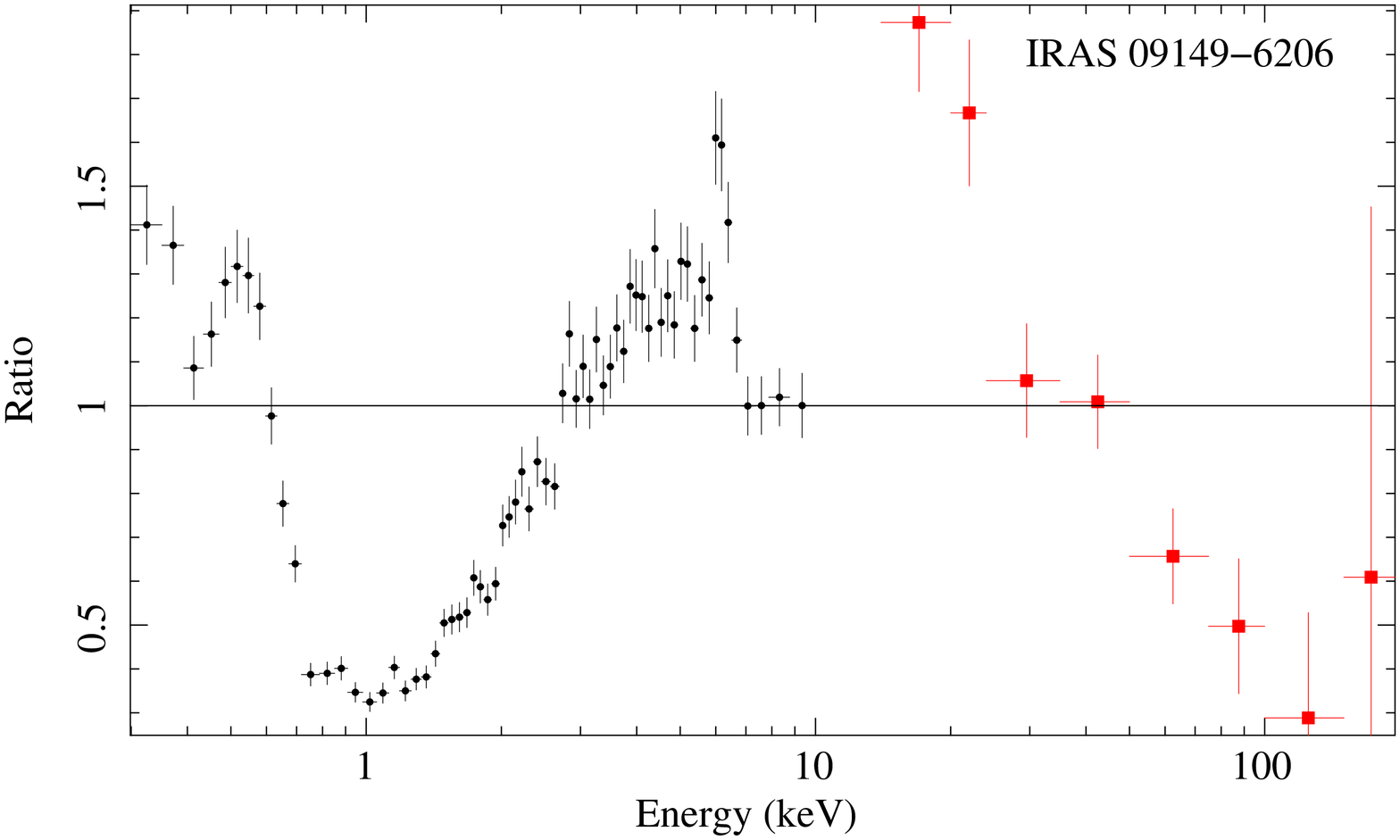}\includegraphics[bb=0bp 0bp 792bp 512bp,scale=0.3]{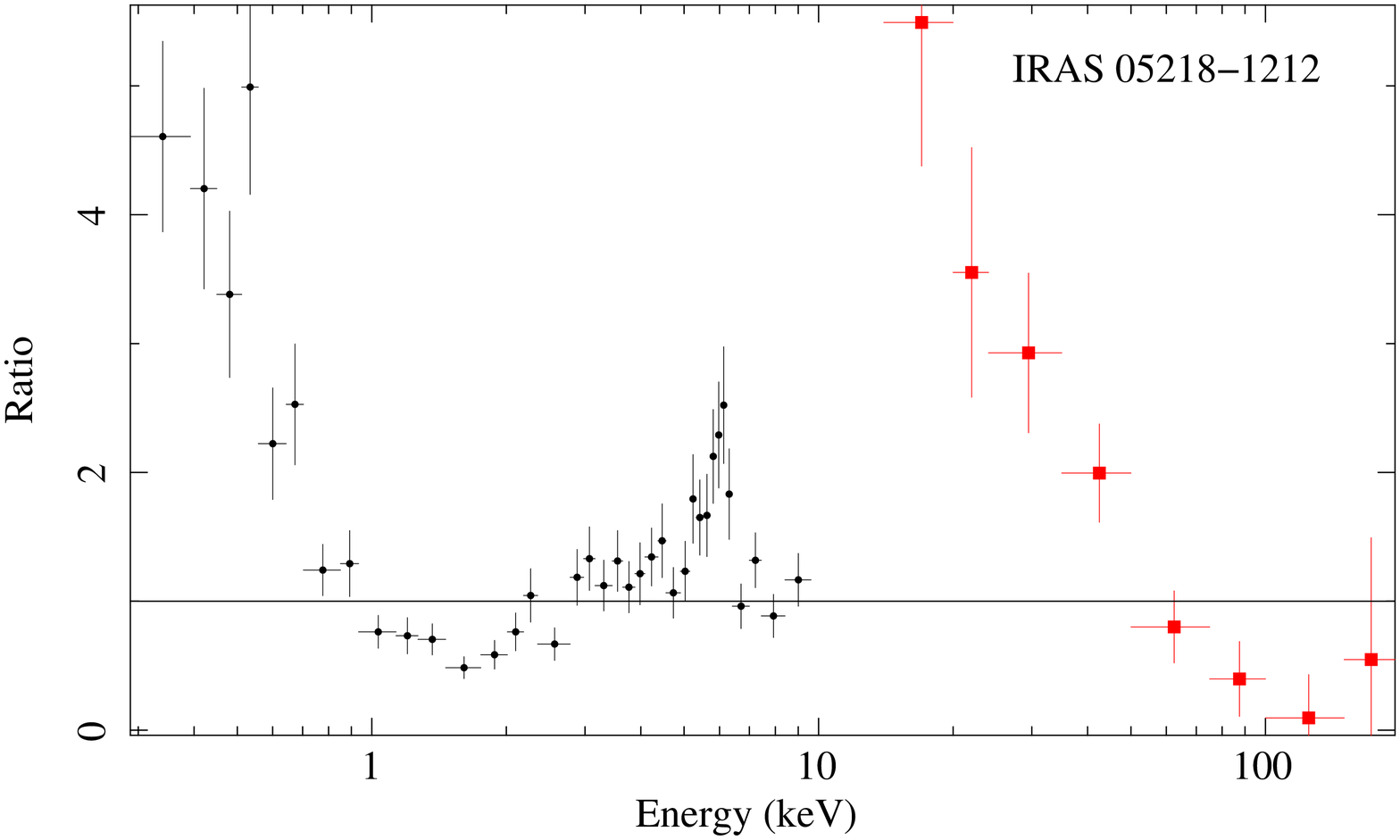}

\caption{\label{fig:ratio plots}The ratio plot of combined \textit{XMM-Newton}  and
\textit{Swift}-Bat spectra of our sample. The model is a simple power law fitted
to the 2.0-10.0 keV band and then extended to other bands. Only EPIC
pn data are plotted for clarity. }

\end{figure}

\begin{figure}[h]

\ContinuedFloat
\caption{(Continued)}
\includegraphics[bb=0bp 0bp 792bp 512bp,scale=0.3]{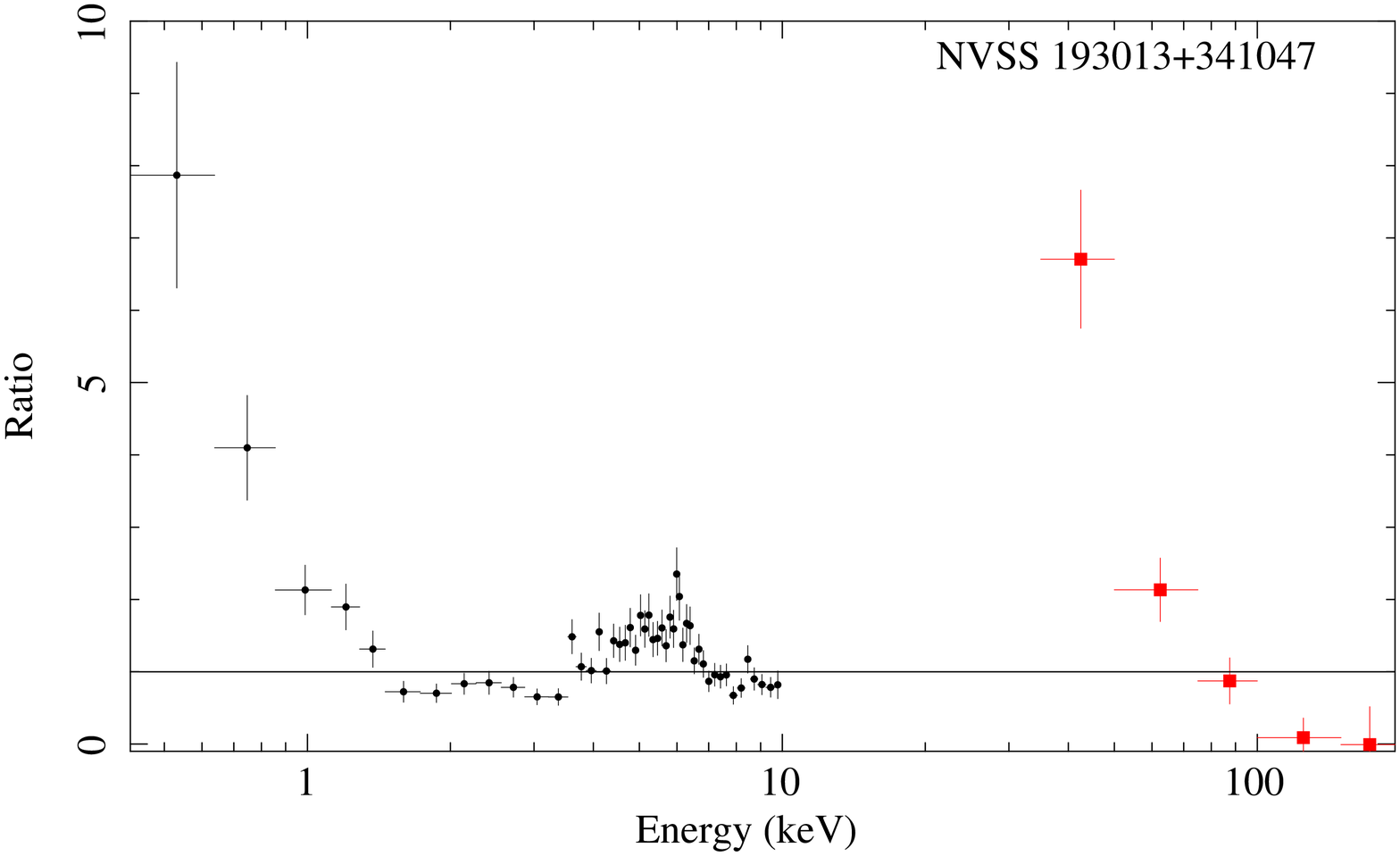}\includegraphics[bb=0bp 0bp 792bp 512bp,scale=0.3]{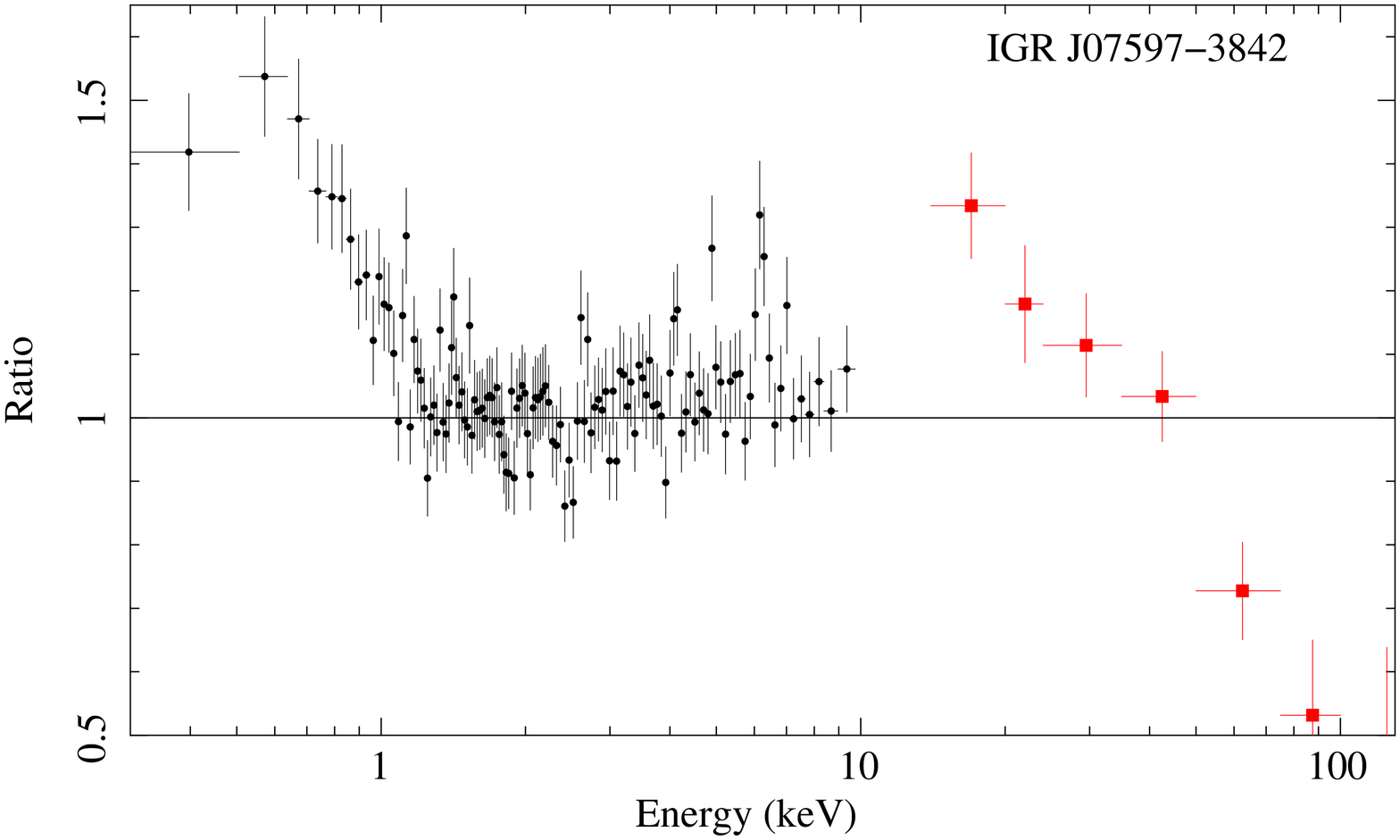}

\includegraphics[bb=0bp 0bp 792bp 512bp,scale=0.3]{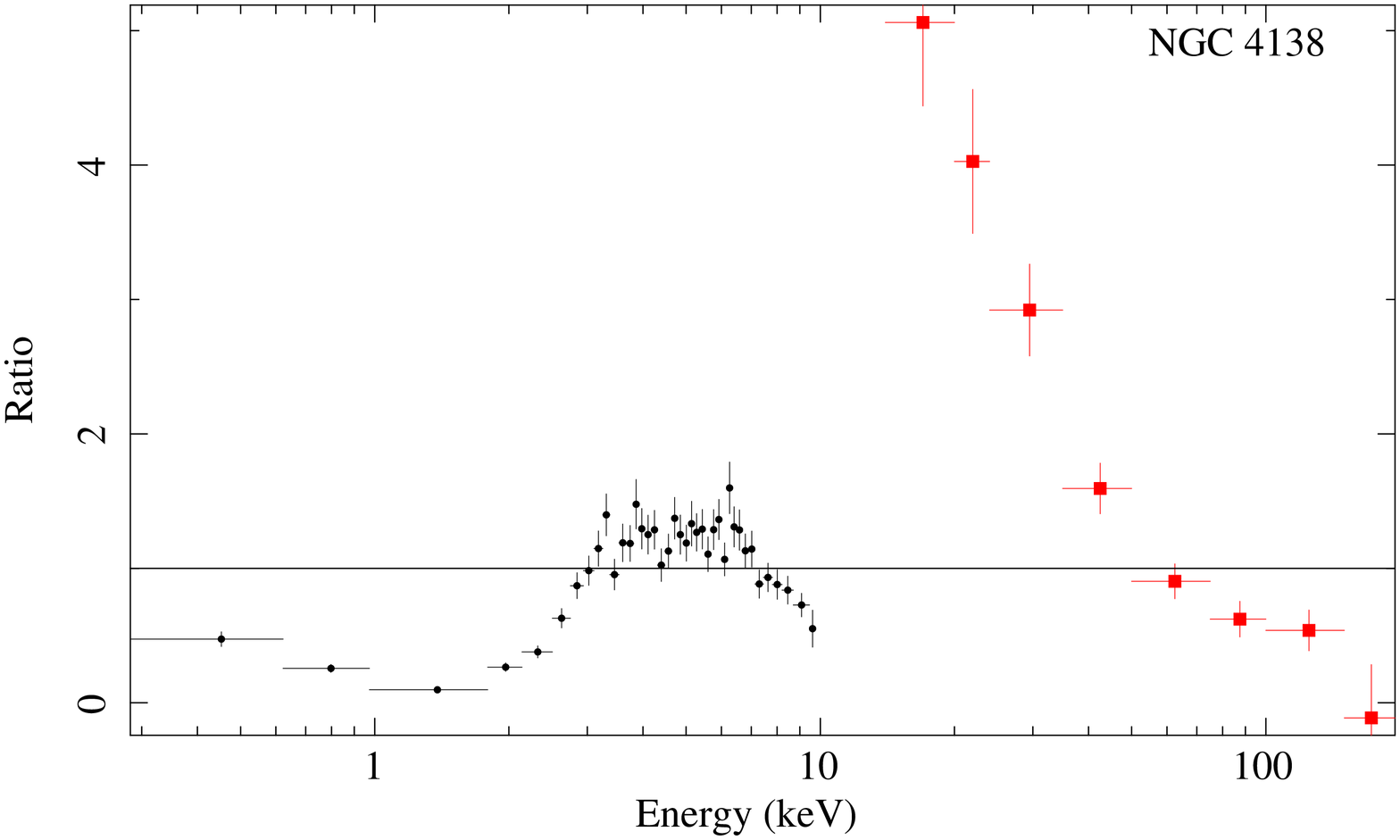}\includegraphics[bb=0bp 0bp 792bp 512bp,scale=0.3]{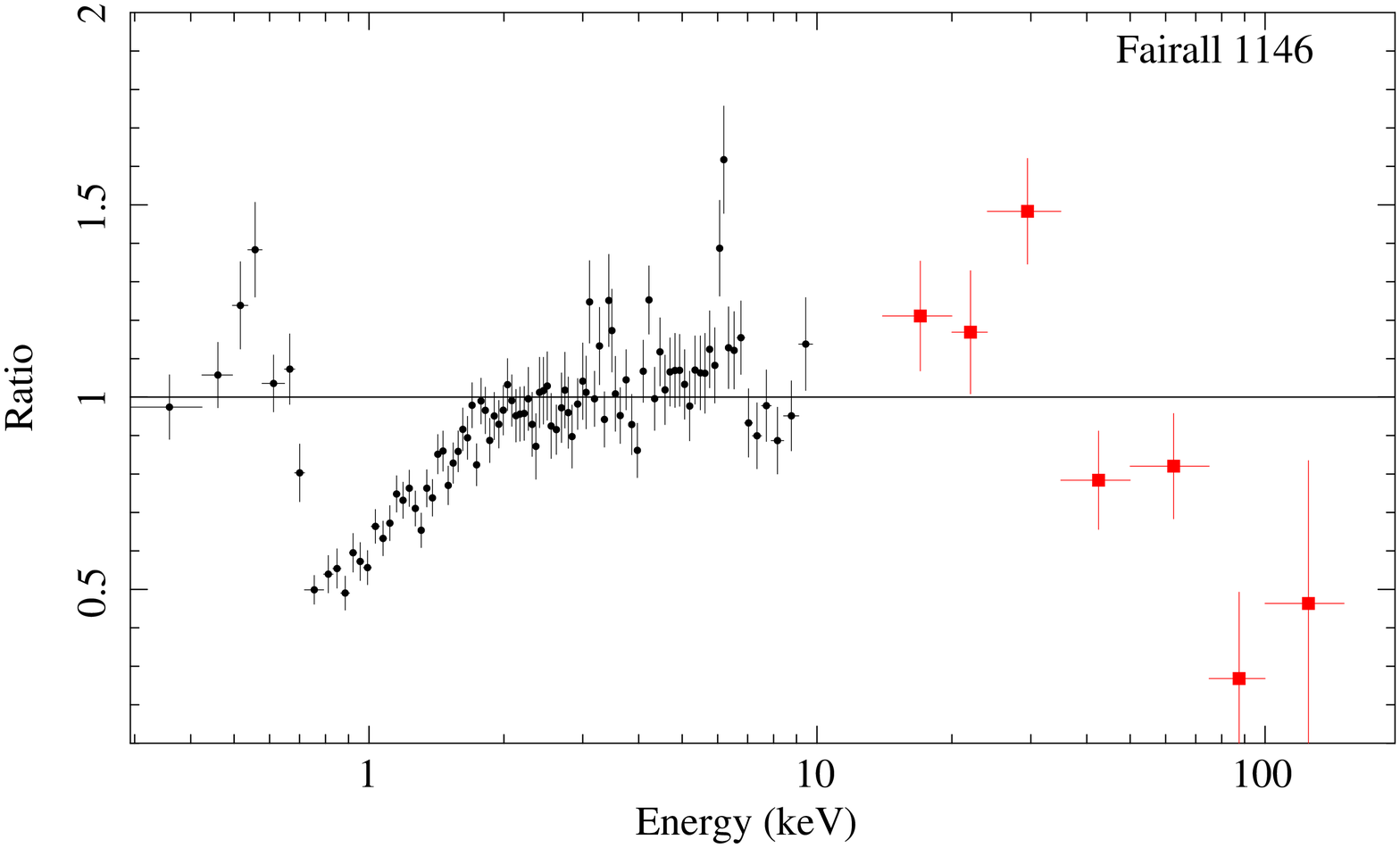}

\includegraphics[bb=-400bp 0bp 792bp 512bp,scale=0.3]{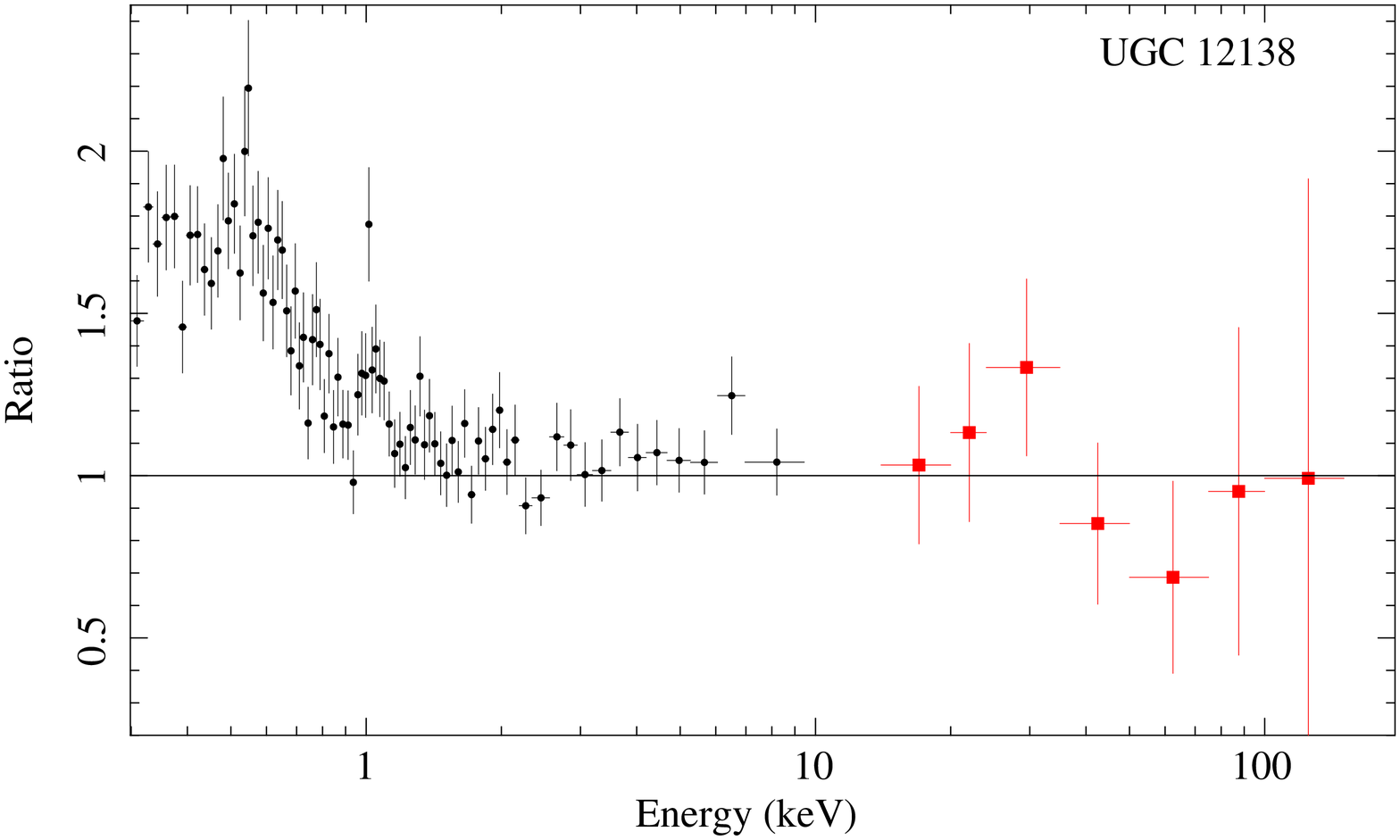}

\end{figure}

\end{document}